\documentclass[reprint,amsmath,amssymb,pre,longbibliography,
  nofootinbib,
]{revtex4-2}
\usepackage{comment}
\usepackage{color}
\usepackage{ulem}
\usepackage{graphicx}
\usepackage{subfigure}
\usepackage{dcolumn}
\usepackage{bm}
\usepackage{multirow}
\usepackage{amsmath}

\usepackage{braket}
\usepackage{diagbox, eqparbox, hhline}
\usepackage{hyperref}
\usepackage{tikz-cd}
\tikzcdset{every label/.append style = {font = \normalsize}}

\DeclareMathAlphabet{\mathpzc}{OT1}{pzc}{m}{it}

\begin{document}

\title{Polarization and dynamic phases of aligning active matter in periodic obstacle arrays}
\author{Daniel Canavello$^1$, C. Reichhardt$^2$, C.J.O  Reichhardt$^2$, and Cl\'ecio C. de Souza Silva$^1$}
\email{clecio.cssilva@ufpe.br}
\affiliation{$^1$ Departamento de Física, Centro de Ciências Exatas e da Natureza, Universidade Federal de Pernambuco, Recife--PE, 50670-901, Brasil}
\affiliation{$^2$ Theoretical Division and Center for Nonlinear Studies, Los Alamos National Laboratory, Los Alamos, New Mexico 87545, USA
}

\begin{abstract}
We numerically examine a system of monodisperse self-propelled particles interacting with each other via simple steric forces and aligning torques moving through a periodic array of obstacles. Without obstacles, this system shows a transition to a polarized or aligned state for critical alignment parameters.
In the presence of obstacles, there is still a polarization transition, but for dense enough arrays, the polarization is locked to the symmetry directions of the substrate.
When the obstacle array is made anisotropic,
at low densities the particles can form a quasi-isotropic state where the system can be polarized in any of the dominant symmetry directions. 
For intermediate anisotropy, 
the particles self-organize into a coherent lane state with one-dimensional polarization. In this phase, a small number of highly packed lanes are adjacent to less dense lanes that have the same polarization, but lanes further away
can have the opposite polarization, so that global polarization is lost. For the highest anisotropy, hopping between lanes is suppressed, and the system forms
uniformly dense uncoupled but polarized lanes.
\end{abstract}

\date{\today}

\maketitle

\section{Introduction}

Active matter systems exhibit self-motility and can arise in biological and soft matter systems \cite{Marchetti2013review, Bechinger2016}.
One of the simplest examples of collectively interacting active matter systems is self-propelled disks
with steric interactions where, at a critical activity level, there is a transition from a fluid to a motility
induced phase-separated state in which a high density solid coexists with a
low density fluid \cite{Fily_2012,Redner_2013,Palacci13,Cates_2015}.
Active matter systems can also be coupled to some form of
periodic substrate \cite{Volpe11,Bechinger2016,Ribeiro20,Reichhardt21-clogging,Kjeldbjerg22,Modica23,reichhardt2023pattern,Chan24, caprini2020activity},
where the motion of single particles can become directionally locked to the substrate symmetry. \cite{Volpe11, BrunCosmeBruny20,Reichhardt20,Nabil22, pattanayak2019enhanced}.
For collectively interacting particle systems in the presence of a periodic substrate, active commensuration effects can arise~\cite{fazelzadeh2023active, nayak2023driven}. Periodic obstacle arrays can screen out social interactions in fish schools leading to a random spread of swimming directions above a critical obstacle density~\cite{ventejou2024behavioral}. When active disks are coupled to a random obstacle array, the activity can generate motility-induced jamming or clogging effects \cite{Reichhardt21-clogging}.

In one variation of active matter particle systems, an additional torque interaction term is present that causes an alignment of the particles when they interact with each other~\cite{bowick2022symmetry} or with confining potentials~\cite{Dauchot2019}. Even at the single-particle level, the presence of aligning torques induced by confining potentials produces a rich phenomenology, such as closed and chaotic orbits and transitions between them~\cite{Dauchot2019, Rubens2022, Rubens2023}. For many interacting active particles, interparticle aligning torques can lead to polarization or flocking effects in which all of the particles move in the same direction \cite{giavazzi2018flocking, geyer2019freezing, caprini2020spontaneous, giavazzi2018flocking, fazli2021active}.
When an active matter system with alignment interactions is subjected to some form of confinement, various types of coherent flows such as vortex-like states can occur \cite{Daniel2024}. In the case of densely packed particles, active crystals exhibiting unconventional collective phenomena emerge~\cite{briand2018spontaneously, baconnier2022selective, baconnier2024self, briand2016crystallization}.
An open question is how active matter systems with alignment interactions behave when interacting with a periodic obstacle array. For example, it is not known whether polarization effects would still occur, or whether the symmetry of the array would cause a symmetry breaking of the possible polarization directions the particles could follow.

In this work, we consider active matter particles with both steric repulsion
and aligning torque effects. In the absence of a substrate, this system shows a transition
to a polarized state at a critical torque alignment. When a periodic square array of obstacles is introduced, the critical value of the alignment parameters remains
unchanged. When the obstacle density is low,
the motion of the particles can
polarize into any direction,
but as the obstacle density increases,
the system becomes polarized along the dominant substrate symmetry directions of $0^{\circ}, 90^{\circ}, 180^{\circ}$, and $270^{\circ}$. 
For increasingly anisotropic obstacle arrays, in which
the anisotropy is controlled by adding more obstacles
along only one direction, a 
transition from two-dimensional (2D) global polarization to
one-dimensional (1D) polarization, or a laned state, occurs.
In the laned state, some of the lanes may be flowing in the direction
opposite to the flow in other lanes.
At intermediate anisotropy,
particles can jump between the lanes
and the system forms coherent lanes in which neighboring lanes have the
same polarization but may have different densities of particles.
At large anisotropy, particle hopping from lane to lane no longer occurs, and individual lanes flow independently; however, the polarization of the individual lanes is still aligned with the anisotropy direction and may be in the
positive $x$ or negative $x$ direction in any given lane.
Our results show that polarization or flocking in active matter with aligning interactions can be controlled by the symmetry of a periodic substrate, and that periodic substrates could thus be used as a new method for guiding and controlling flows in aligning active matter systems.

\section{Model details}

Our system consists of $N$ active Brownian particles (ABP) and $N_{\text{p}}$ obstacles, or posts, confined in a 2D plane. The $i$-th particle has position and orientation given by $\bm{r}_i = (x_i, y_i)$ and $\hat{\bm{n}}_i = (\sin \theta_i, \cos \theta_i)$ respectively. The posts are modeled as immobile particles.

The translational and orientational dynamics of the active particles are modeled by the following overdamped equations of motion (see Ref.~\cite{Daniel2024} for more details):
\begin{align}
  \label{eq:motion}
  \dot{\bm{r}}_i &= v_0 \hat{\bm{n}}_i + \mu\bm{F}_i, \\
  \dot{\theta}_i &= \beta (\hat{\bm{n}}_i \times {\bm{F}}_i)\cdot \hat{\bm{z}} + \sqrt{2D}\zeta_i(t),
  \label{eq:orientation-part}
\end{align}
where $v_0$ denotes the propulsion speed, $\hat{\bm{z}}$ is the unit vector perpendicular to the $xy$ plane, $\mu$ ($\beta$) is the translational (angular) mobility, and $\bm{F}_i$ is the total conservative force acting on active particle $i$ due to neighboring active particles and obstacles. The latter interaction is modeled by a purely repulsive short-range Weeks-Chandler-Anderson (WCA) potential: $U(r)=4\epsilon[(\sigma/r)^{12}-(\sigma/r)^6]-\epsilon$ for $r<2^{1/6}\sigma$, and $U(r)=0$ for $r>2^{1/6}\sigma$, where $\sigma$ is the particle size and $\epsilon$ is the interaction energy scale.
The noise term satisfies $\langle \zeta_i(t) \zeta_j(t') \rangle = \delta_{ij}\delta(t - t')$, with $D$ representing the rotational diffusion constant~\cite{Howse_2007,volpe2011,Drescher2011}.
The first term on the right-hand side of Eq.~\eqref{eq:orientation-part} gives the restoring torque produced by the total force $\bm{F}_i$ acting on particle $i$.
Similar torque terms have been used to model collective phenomena in biological clusters, dense arrays of soft particles, and vibrating disks~\cite{Shimoyama1996,Henkes2011,Weber2013,Lam2015}.

In our work, the interparticle aligning torques produce unique phenomena that could not be observed otherwise, such as the full polarization of the system even at low particle densities. This orientational ordering of the particle motion arises from multiple collisions among the particles. In each collision, the particles exert torques on one another that act to partially align their individual polarizations, which can lead to the progressive ordering of the whole system in velocity space.  The parameter $\beta$ controls how fast this alignment happens (that is, how large the angle difference will be after a collision). For large $\beta$ only a small number of collisions is needed to produce full polarization, leading to the rapid alignment of the system. For small $\beta$, the small correlations acquired during each collision are more easily destroyed by noise, and in this regime it is possible that full alignment might never happen. 

We consider a square simulation box of side $L=50\sigma$ with periodic boundary conditions. In the case where obstacles are present, we focus on two kinds of periodic obstacle arrays: square arrays containing the same number of posts in both directions, and rectangular or anisotropic arrays where the number of rows of obstacles is fixed while the number of columns is increased, thus forming horizontal channels. 
We note that when there are 25 obstacles per row, the horizontal distance between posts is $2\sigma$ and the free space not occupied by obstacles is equal to $\sigma$, the particle diameter. For this obstacle configuration, particles rarely switch lanes, and when there are 26 or more obstacles per row, lane switching becomes impossible. 

\section{Results}

\subsection{Collective behavior without obstacles} \label{sec:NoPosts}

\begin{figure}[t]
  \includegraphics[width= \linewidth]{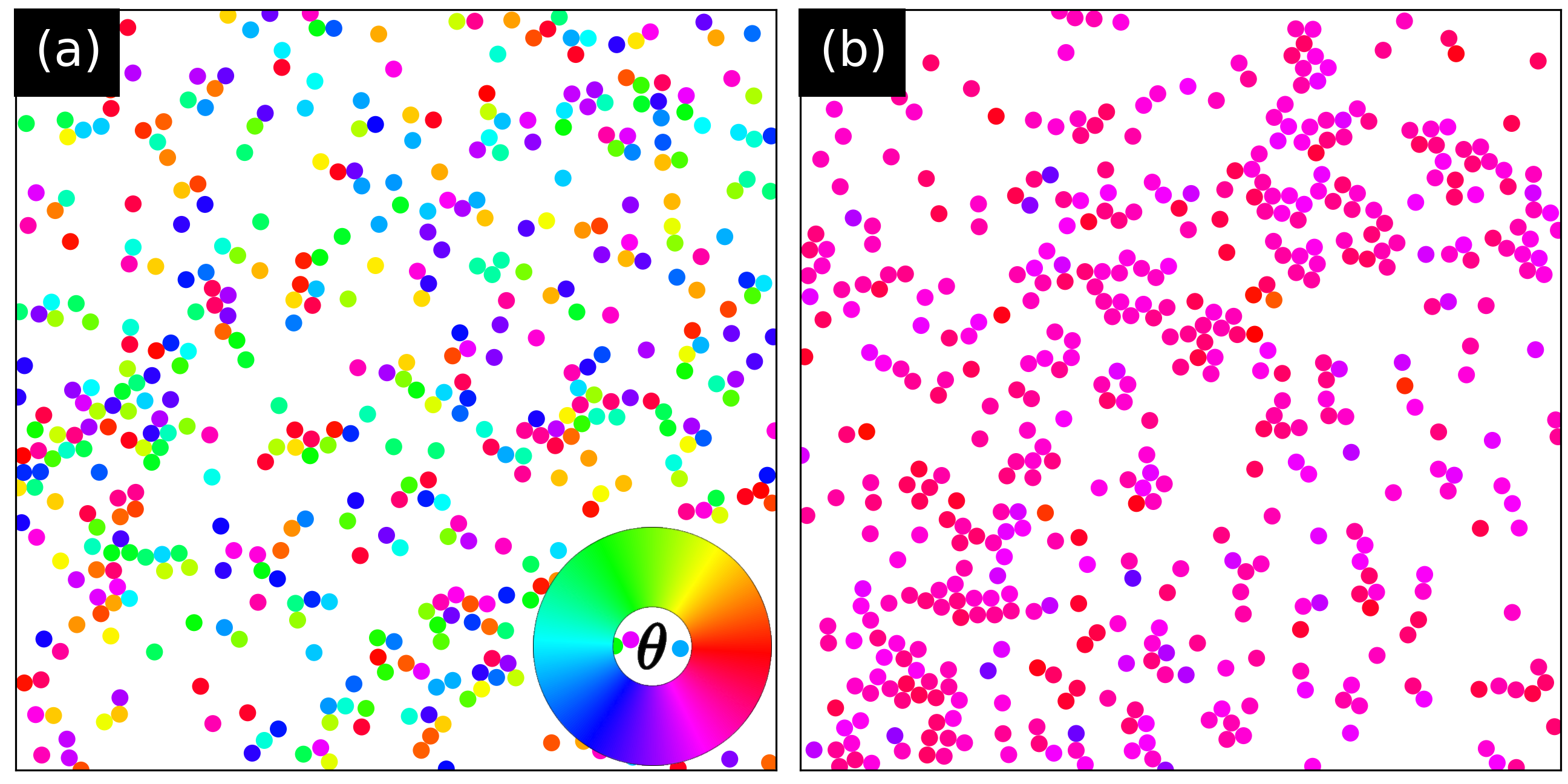}
  \caption{Images of the particle locations and polarization $\theta$ for a system without obstacles
    with $N = 480$ particles at $D = 0.01$. (a) $\beta = 0.1$ in the unpolarized state. (b)  $\beta = 0.6$ in the polarized state.
  }
  \label{fig:1}
\end{figure}

In the absence of obstacles and in the presence of finite noise, we observe that the particles move randomly as a gas for small angular mobility $\beta$, exhibiting only a small average velocity.
Above a finite critical value $\beta>\beta_c$, a phase transition takes place
to a state with a significant amount of polarization vector alignment,
causing the motion to occur
predominantly in a single direction and producing large average velocity values.
In Fig.~\ref{fig:1}(a) we show a non-polar state
in a system with $N=480$ at $\beta/D = 10$,
while in Fig.~\ref{fig:1}(b) we illustrate the same system in the polarized state at  $\beta/D = 60$ where all particle motion is generally in the same direction. Here and in what follows, $\beta$ and $D$ are expressed in units of $\mu/\sigma$ and $v_0/\sigma$, respectively.
%
The polarization transition can be understood by noticing that the collective dynamics are ruled by two characteristic times:
(i) the polarization time $\tau_P$, which is the typical time necessary to fully polarize the system through collisions from a random configuration,
and (ii) the persistence time $\tau_D=1/D$, which is a measure of the time a particle can travel freely without significantly changing its orientation due to thermal fluctuations.
The polarization time can be roughly estimated as $\tau_P\sim\mu L^2/(\beta N\sigma^2 v_0)$.
\footnote{
  The polarization time can be expressed as $\tau_P \sim N_\text{col}\tau_{\text{col}}$, where $N_\text{col}$ is the typical number of collisions necessary to fully align the orientation of a particle with the mean polarization, and $\tau_{\text{col}}=1/(\sigma n v_0)=L^2/(\sigma N v_0)$ is the collision time. The deflection $\delta\theta$ induced by the interparticle torques during each collision brings the particle orientation ever closer to the average polarization. Therefore, assuming the particle is initially misoriented, $N_\text{col}\sim 1/\delta\theta$. From Eq.~\eqref{eq:orientation-part} and noting that during a collision, the interaction force is of order $v_0/\mu$, $\delta\theta$ can be estimated as $\delta\theta \sim \beta(v_0/\mu)\delta t \sim \beta\sigma/\mu$. Therefore, $\tau_P\sim\mu L^2/(\beta N\sigma^2 v_0)$.
}

\begin{figure}[t]
  \includegraphics[width= \linewidth]{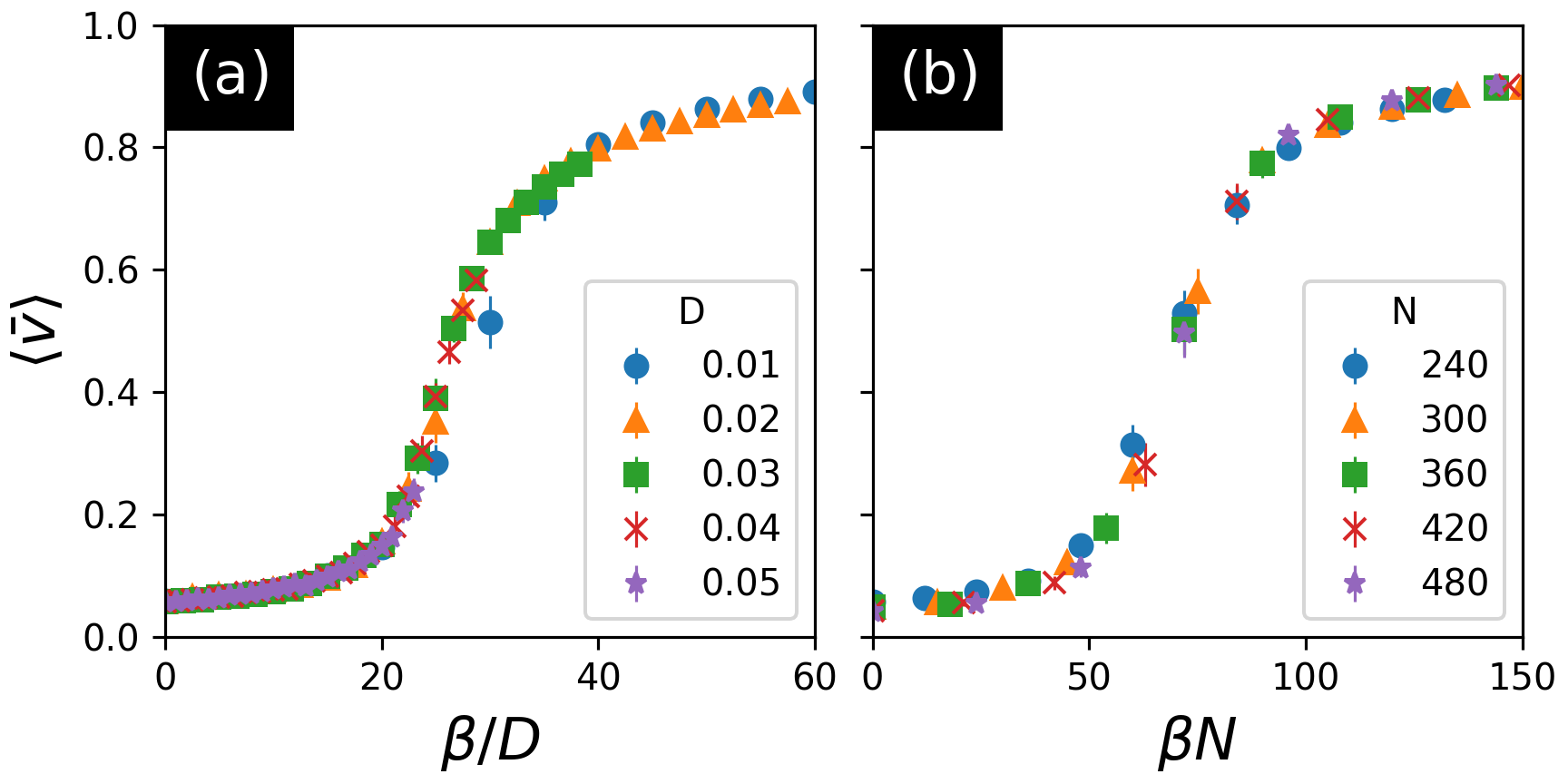}
  \caption{(a) The system polarization $\langle \bar{v}\rangle$ vs $\beta/D$ for $D=0.01,$ 0.02, 0.03, 0.04, and $0.05$ in samples with $N = 240$ showing a transition from the unpolarized state to a polarized state near $\beta/D = 25$.
  (b) $\langle \bar{v}\rangle$  vs $\beta N$ for $N = 240, 300, 360, 420$ and $480$ in samples with $D = 0.01$ showing that the polarization transition occurs near $\beta N = 70$. The error bars are standard deviations from the mean obtained from 100 independent realizations of the random noise.
  }
  \label{fig.2}
\end{figure}

Since full polarization is only possible when $\tau_P<\tau_D$, we can estimate that the transition from the unpolarized to the polarized state should occur when $\beta_c \sim\mu D/(v_0f)$, where $f=N\sigma^2/L^2$ is the filling fraction.
To characterize the orientational ordering of the system, we compute an average of the center-of-mass (cm) velocity over time and random noise, $\bar{\bm{v}}=\frac{1}{N}\sum_i\bm{v}_i$, for different values of $\beta$, $N$, and $D$.
Scaling $\beta$ with $D$ and $f$ according to the simple estimate above, we obtain a collapse of the simulation data, as shown in Fig.~\ref{fig.2}(a) where we plot $\langle \bar{v}\rangle$ versus $\beta/D$ at $D = 0.01, 0,02, 0,03, 0,04$ and $0.05$ for $N = 240$, where the transition occurs near $\beta/D = 25$.
A similar polarization transition was also observed in Ref.~\cite{Lam2015} for a model of interacting polar disks where the aligning torques were modeled as $\hat{\bm n}_i\times{\bm v}_i/|{\bm v}_i|$, different from the quantity $\hat{\bm n}_i\times\bm{F}_i$ that we use in our model. In spite of this difference in the models, we anticipate that the same scaling laws we find should also be observable in the model from Ref.~\cite{Lam2015} in the overdamped limit, since this follows straightforwardly from a time-scale analysis similar to the one described above. 
Figure~\ref{fig.2}(b) shows $\langle \bar{v}\rangle$ versus $\beta N$ at fixed $D = 0.01$ for  $N = 240$ to $N = 480$ particles corresponding to filling fractions $f=N\sigma^2/L^2=0.096$ to $0.192$.
These values are considerably below the threshold for the observation of MIPS in the absence of aligning torques, $f\gtrsim0.5$ \cite{Roca2021}.
We find that the polarization transition occurs near $\beta N = 70$.

\subsection{Effect of a square array of obstacles}
\label{sec:SquareArray}

\begin{figure}[!t]
\includegraphics[width= 0.8\linewidth]{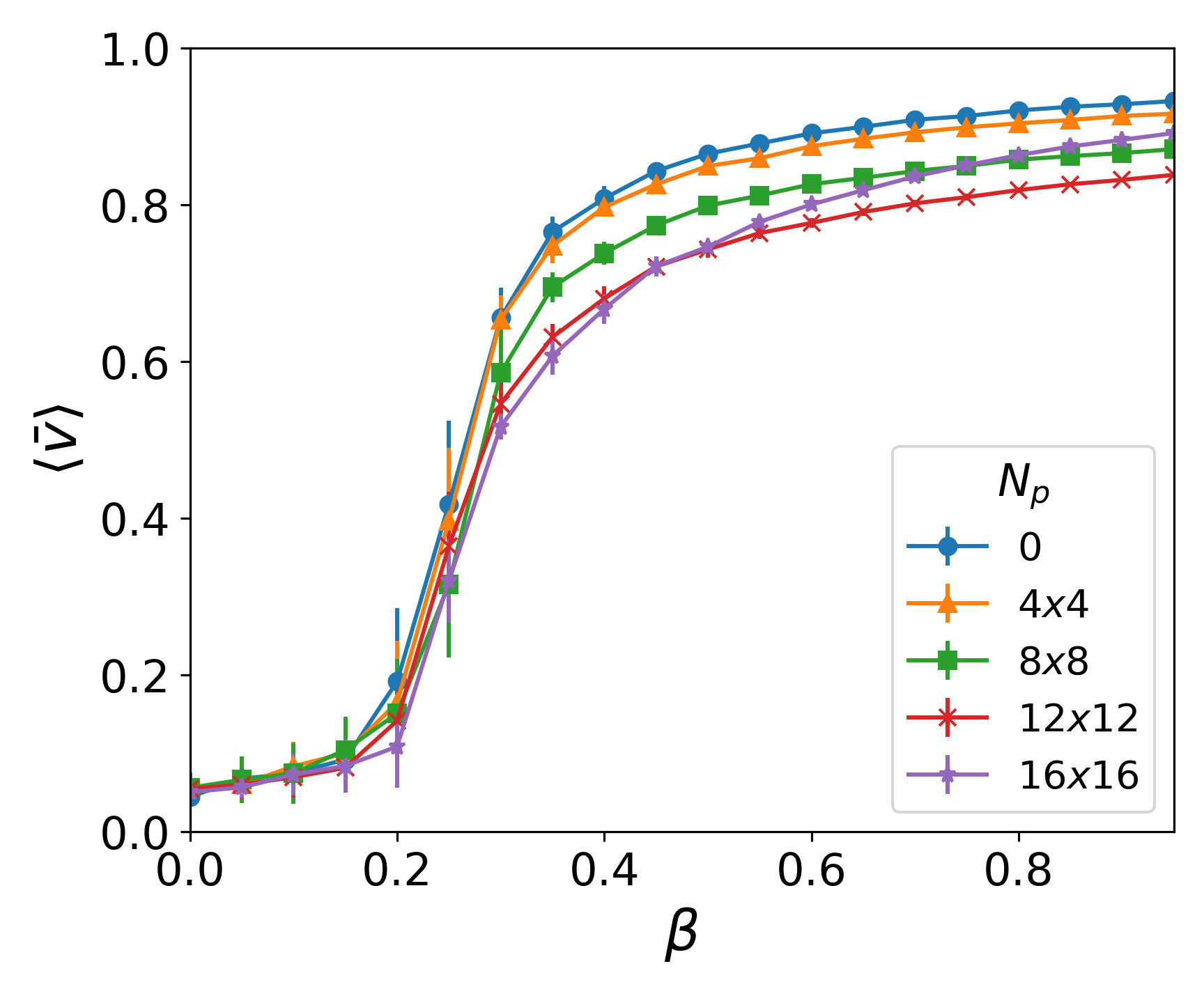}
\caption{The system polarization $\langle \bar{v}\rangle$
vs $\beta$ for the system from Fig.~\ref{fig.3} with square obstacle arrays
at $D = 0.01$ and $N = 240$ for no obstacles,
$4 \times 4$ obstacles, $8 \times 8$ obstacles,
$12 \times 12$ obstacles, and
$16 \times 16$ obstacles. The value of $\beta$ at which the
polarization occurs does not change as the obstacle density varies.
The error bars indicate the standard deviation from the mean for 50 independent realizations of the random noise.
}
\label{fig.4}
\end{figure}

We next consider a system with a square array of obstacles. We find that adding obstacles does not change the parameters at which the polarization transition occurs.
This is illustrated in Fig.~\ref{fig.4}, where we plot the
system polarization versus $\beta $ for $N_{\text{p}} = 0, 4 \times 4, 8 \times 8, 12 \times 12 \text{ and } 16 \times 16$ in samples with $D=0.01$ and $N=240$.
Here, the transition to the polarized state
occurs near $\beta = 0.3$, with a negligible
shift to higher $\beta$ appearing for increasing obstacle density. However, the obstacle array strongly influences  the directions along which the polarization can occur.

\begin{figure}[!t]
\includegraphics[width= \linewidth]{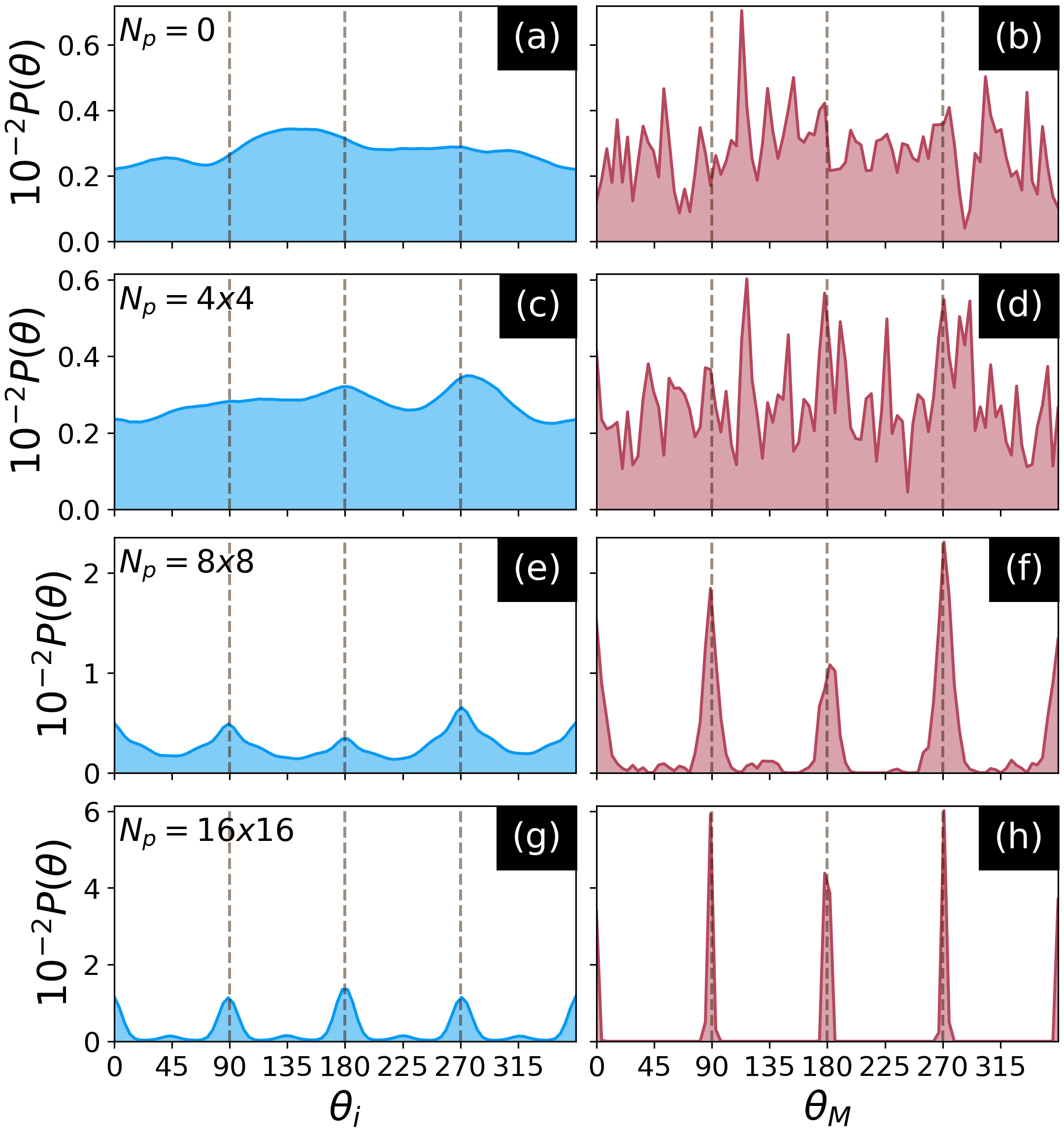}
\caption{(a,c,e,g) Normalized histograms of the distribution
  $P(\theta_i$)
  of the angular velocity of the individual particles.
  (b,d,f,h) Normalized histograms of the distribution
  $P(\theta_M)$ of the mean direction of motion of the whole system.
  Here $N = 240$, $\beta =1.0$, $D = 0.01$, and the histograms are accumulated over 360 disorder realizations.
  The dashed lines indicate the symmetry directions of the square array at 0$^\circ$, 90$^\circ$, 180$^\circ$, and 270$^\circ$.
  (a,b) No obstacles showing a lack of directional locking.
  (c,d) $4\times4$ obstacles showing only slight locking.
  (e,f) For $8\times8$ obstacles, both the individual and mean velocities begin  to lock to the substrate symmetry directions.
  (g,h) For $16\times 16$ obstacles, the mean motion occurs only along the symmetry directions of the substrate.
}
\label{fig.3}
\end{figure}

In Fig.~\ref{fig.3}(a,c,e,g) we plot normalized histograms of the distribution of individual polarization angles $P(\theta_i)$ averaged over 360 independent realizations of the random noise starting from different initial conditions.
Here $N =240$, $\beta=1.0$ and $D=0.01$, so the system is in a polarized regime. 
Figure~\ref{fig.3}(b,d,f,h) shows the corresponding histograms of the distribution of the angles of the total polarization vector $\bm{M}=\frac{1}{N}\sum_i\hat{\bm n}_i$.
When no obstacles are present, that is $N_{\text{p}} = 0$, Fig.~\ref{fig.3}(a,b) shows that there are no favored directions for motion.
For $N_{\text{p}} = 4 \times 4$, in Fig.~\ref{fig.3}(c,d) there is very weak locking along some of the symmetry directions, while for $N_{\text{p}} = 8 \times 8$, Fig.~\ref{fig.3}(e,f) shows that there is pronounced locking of both the
velocity and the net polarization along the four symmetry directions.
As the number of obstacles increases, the locking effect becomes sharper, as seen for the $N_{\text{p}} = 16\times 16$ array in Fig.~\ref{fig.3}(g,h).
For the $16 \times 16$ obstacle array, the mean velocity direction shown in Fig.~\ref{fig.3}(h) is always aligned with multiples of $90^{\circ}$, with zero weight at other angles;
however, the individual particle velocities in Fig.~\ref{fig.3}(h) exhibit small additional peaks near multiples of $45^{\circ}$ due to the fact that some particles can switch lanes between the obstacles.
This result indicates that when the square obstacle array is sufficiently dense,
the polarization becomes locked to the dominant symmetry directions of the substrate lattice.

The directional locking we observe is similar to that found for run and tumble particles
with long run lengths on square arrays where, during the run phase, the motion becomes locked to certain array symmetry directions
\cite{Volpe11, BrunCosmeBruny20, Reichhardt20}. In the polarized active matter case, the directional locking arises from
a combination of the collective interactions, which create
the velocity alignment, and symmetry breaking by the substrate.
We expect the polarization to be locked to multiples of $60^{\circ}$ for a triangular substrate. 

\subsection{Rectangular arrays: laning transition}

\begin{figure*}[!t]
\includegraphics[width= \linewidth]{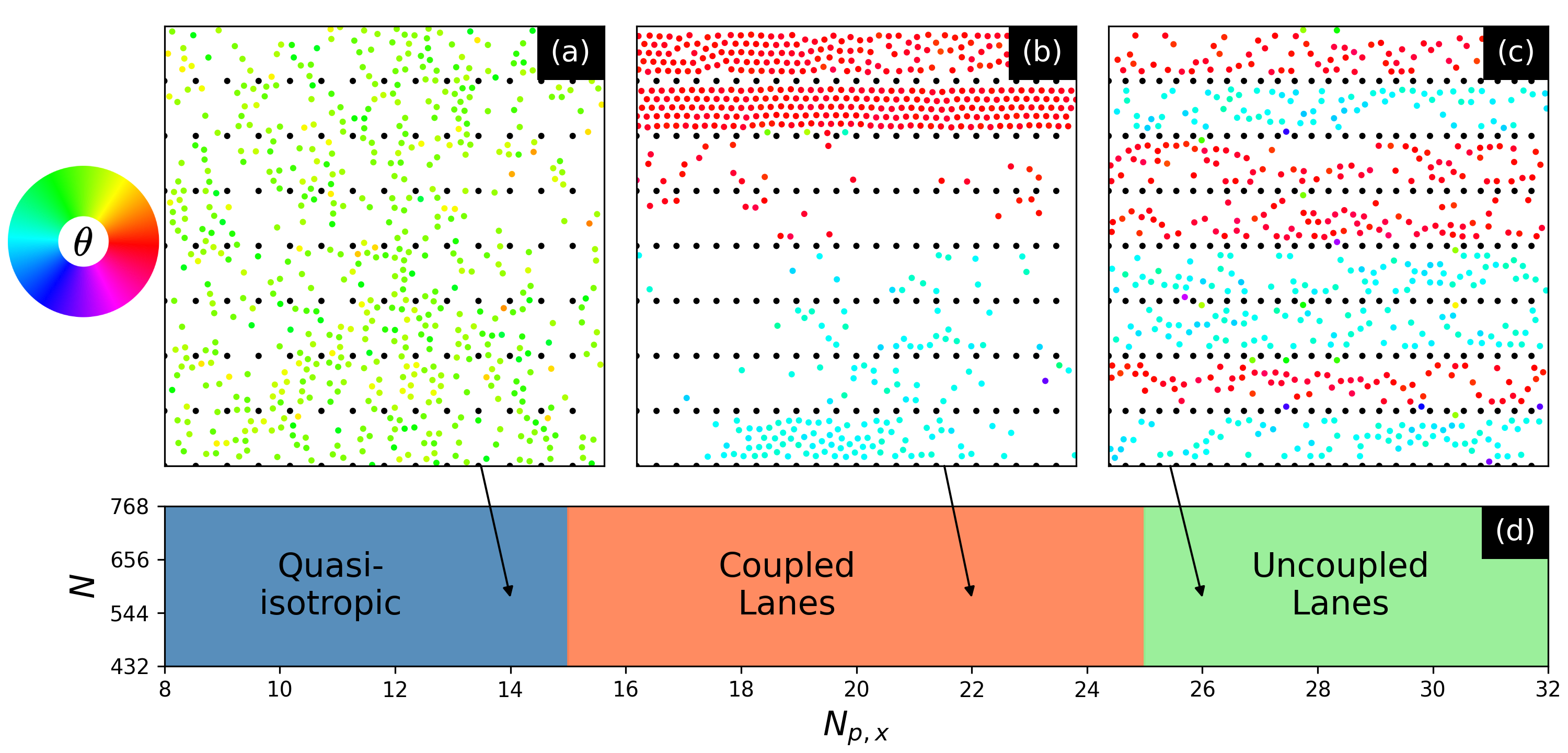}
\caption{
(a,b,c) Images of the particles and (d) a schematic phase diagram
  for a system with an anisotropic obstacle array where the
  number of obstacles along the $x$ direction $N_{\text{p}, x}$ is increased
  while the number of obstacles along the $y$ direction $N_{\text{p}, y}$ remains fixed
  at 8.
  Here $\beta = 5.0$, $D = 0.01$ and $N = 576$.
  The particle color indicates the particle polarization direction
  $\theta_i$ as shown by the color wheel,
  and the obstacles  or posts appear as black circles.
  (a) The quasi-isotropic state in a system with $N_{\text{p}, x}=14$,
  where the particle polarization can be oriented in any of the
  four major symmetry directions.
  (b) A coupled lanes state in a system with $N_{\text{p}, x}=22$,
  where the polarization is restricted to lie one direction,
  lane-to-lane hopping can occur, and the system organizes
  into a state with a small number of dense lanes.
  (c) The uncoupled lanes state in a
  system with $N_{\text{p}, x}=27$, where particles
  cannot move between lanes and each lane has the same density.
  (d) Schematic phase diagram as a function of number of particles $N$
  vs $N_{\text{p}, x}$ showing the window spanned by each of the three phases.
}
\label{fig.5}
\end{figure*}

Next, we make the obstacle array anisotropic by increasing the number of obstacles along the $x$ direction, denoted by $N_{\text{p}, x}$, while holding the number of obstacles along the $y$ direction fixed to $N_{\text{p}, y} = 8$, so that the ratio of the lattice constant in the $x$ direction to that in the $y$ direction changes over the range 1:1 to 4:1.
In Fig.~\ref{fig.5}, images and a schematic phase diagram illustrate the behavior of a system in which we start with an $ 8\times8$ obstacle array and gradually increase $N_{\text{p}, x}$.
The color of each particle indicates its polarization direction, and the obstacles are shown as black circles. 
Here, red indicates motion in the positive $x$ direction, blue is for motion in the negative $x$ direction, and green is for motion along the positive $y$ direction.
The images are obtained in samples with $N=576$, $\beta = 5.0,$ and $D =0.01$,
so that the system would be deep in the polarized phase in the absence of the
obstacles.
The first of the three phases we observe
is a quasi-isotropic state, where the polarization can be oriented along
any of the four dominant directions of the substrate,
as shown in Fig.~\ref{fig.5}(a) for a system with $N_{\text{p}, x}=14$.
As the anisotropy increases,
there is a transition from the 2D global polarization to
quasi-1D coupled lanes in which the system forms a series of
lanes that are polarized in either the $+x$ or $-x$ direction.
Although the particles predominantly move along $x$, hopping of the
particles from one lane to the next can occur,
and over time we find that the system organizes into a heterogeneous
lane state in which a few lanes contain a dense assembly of particles while
other lanes contain very few particles.
Nearest neighbor dense lanes are polarized in the same direction, but lanes
that are a greater distance away may have the opposite polarization,
as shown in Fig.~\ref{fig.5}(b) for a system
with $N_{\text{p}, x}=22$.
Since some of the lanes have opposite polarization, the polarization
is not global, unlike the global polarization that we observed in the
quasi-isotropic phase.
For even higher anisotropy, the spacing between the obstacles becomes
smaller than the diameter of the particles, so hopping of particles
between lanes is no longer possible and adjacent lanes become decoupled.
The individual lanes still polarize,
as shown in Fig.~\ref{fig.5}(c) for
a system with $N_{\text{p}, x}=27$, but
the particle clustering observed in the coupled lane state is absent.
Figure~\ref{fig.5}(d) shows a schematic phase diagram as a function of
number of particles $N$ versus
$N_{\text{p}, x}$ indicating where the three regimes occur.

In the ESI, we present three videos that illustrate each phase of the system. For all videos the parameters are set to $\beta = 5.0, N = 576 \text{ and } D=10^{-2}$. 
In the first video, we set $N_{\text{p}, x} = 10$ and the system settles into a downward motion after the initial transient time.
The second video shows $N_{\text{p}, x} = 18$, where the system exhibits coupled lanes. Three densely packed lanes are seen, with sparse lanes separating them. Stray particles can be seen moving vertically until they are captured by dense flocks.
The final video features a system with $N_{\text{p}, x} = 26$, where lanes can no longer communicate with each other. The video is sped up to double time, as the system takes much longer to reach equilibrium. In this phase, we observe a jammed lane that forms and then breaks down due to thermal fluctuations. The spacing between the obstacles is just slightly smaller than the particle size, causing several particles to become stuck in the ridges.

\begin{figure}[!t]
\includegraphics[width= \linewidth]{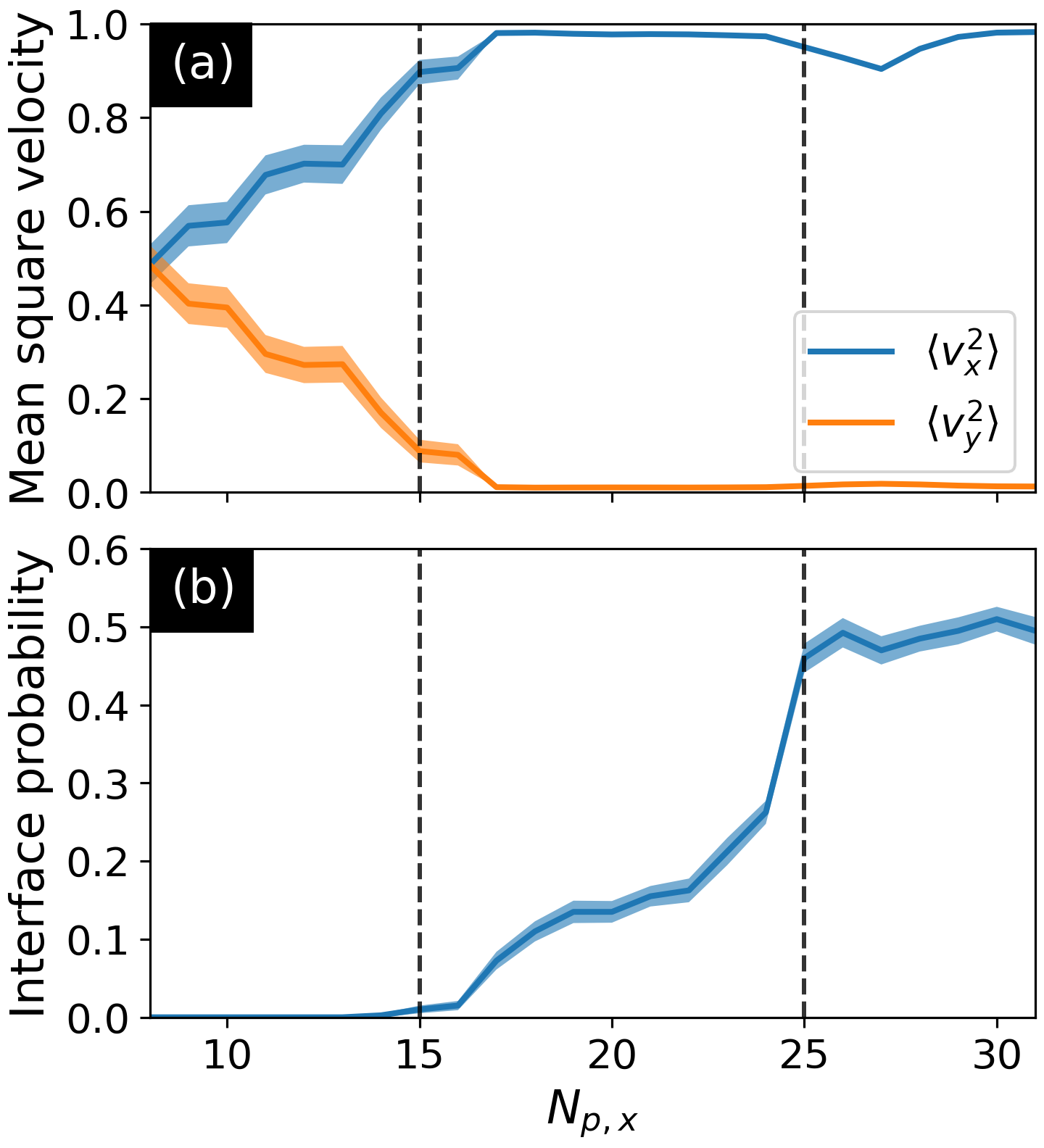}
\caption{
  (a) The mean square velocities $\langle v^2_{x} \rangle$ and $\langle v^2_{y} \rangle$  versus
  $N_{\text{p}, x}$ for systems with $N=576$, $\beta=5.0$, and $D=0.01$.
  (b) The probability that two neighboring lanes have opposite polarities. This quantity is close to $0.2$ for the coupled lanes and increases to $0.5$, indicating random polarization from lane to lane, in the decoupled lane state. The shaded error bands are standard deviations from the mean obtained from 100 independent realizations of the random noise.
}
\label{fig.6}
\end{figure}

To better characterize the different phases, in Fig.~\ref{fig.6}(a)
we plot the mean square velocity $\langle V^2_{x} \rangle$ and
$\langle V^2_{y} \rangle$  versus $N_{\text{p}, x}$.
For $N_{\text{p}, x}=8$ to $N_{\text{p}, x} = 15$,
corresponding to an anisotropy ratio of $1.0$ to $2.0$,
the average velocity is finite in both directions, but the $x$ direction
velocity increases with respect to the $y$ direction velocity as the
anisotropy increases.
For $N_{\text{p}, x}$ greater than $16$, corresponding to
an anisotropy ratio greater than $2.0$, the velocities are almost
exclusively in the $x$-direction since the velocity becomes fully locked
to the anisotropic channels.
In Fig.~\ref{fig.6}(b), we plot the probability that a velocity
interface will appear between adjacent lanes, indicating that the two
lanes are polarized in opposite directions.
In the quasi-isotropic phase,
this quantity is zero since the system
exhibits a global polarization.
For $15 < N_{\text{p}, x} < 25$  where the system is the coupled
lane state, the particles form a small number of highly packed lanes and
neighboring lanes tend to have the same polarization direction,
so the probability of observing a velocity interface is a small
value
close to $0.2$.
When $N_{\text{p}, x}>25$, particles no longer jump between lanes, and the system forms the decoupled lane state in which the velocity interface probability 
is close to $0.5$, indicating random choice of direction for each lane.

In the isolated lane state, the system can become clogged or
jammed when initialized, so a small noise is
required to ``force'' a lane into picking left or right as its
direction of flow.
The transient time required for the system to settle into a steady flow state
decreases with increasing $N$ and increasing $\beta$,
which can be viewed as resulting when each lane
develops impenetrable walls in the vertical direction while maintaining
periodic boundaries in the horizontal direction,
turning each individual lane into a quasi-1D version of the
obstacle-free system analyzed in Section~\ref{sec:NoPosts}.
At the transition from coupled to uncoupled lanes, there is a small drop
in $\langle V^2_{x} \rangle$ that arises when a small number of
particles become stuck on the walls, unable to pass through, causing
a clogging of the traffic in their lane.
As $N_{\text{p}, x}$ increases, the nooks in the walls where the particles
can become stuck shrink in size, leading to smoother flow of the
particles.

\begin{figure}[!t]
\includegraphics[width= \linewidth]{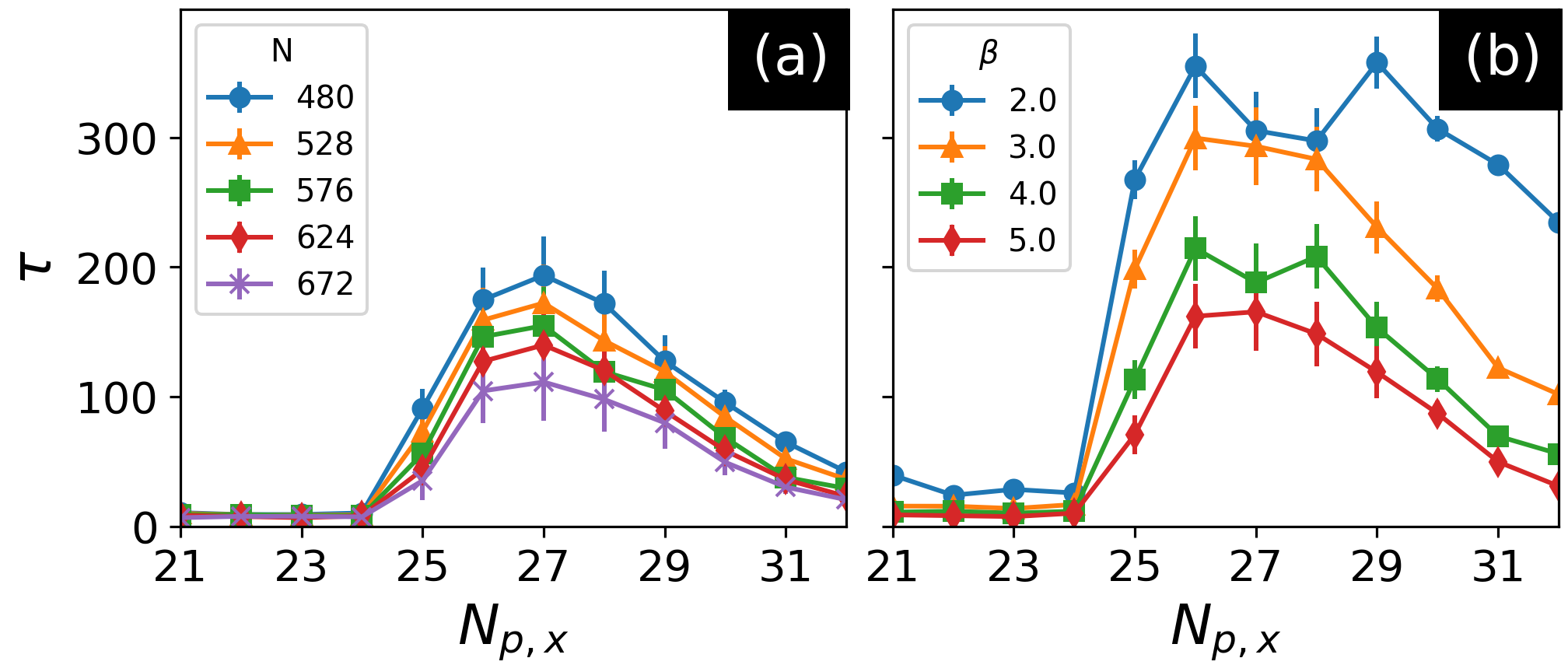}
\caption{
(a) Transient time $\tau$ to reach a steady velocity starting from a random
initial condition vs $N_{\text{p}, x}$
for different values of $N$ at $\beta = 5.0$ and $D=0.01$.
Error bars indicate standard deviations from the mean for 50 realizations
of disorder.
The vertical mobility of particles is small when $N_{\text{p}, x}=25$,
leading to the formation of
jammed states in some of the lanes. These jammed states are gradually
destroyed by noise, but if they occur, $\tau$ increases noticeably.
For higher values of $N$, there are more collisions between particles, and this enables the system to reach a steady state faster.
Similar behavior appears for fixed $N$ when $\beta$ is
increased, as shown in the plot of $\tau$ vs $N_{\text{p}, x}$
in panel (b) for $N=528$.
}
\label{fig.7}
\end{figure}

In Fig.~\ref{fig.7}(a) we show the dependence of the mean transient time
$\tau$ to reach steady flow from a random initial configuration
on $N_{\text{p}, x}$ for
different values of $N$
in the decoupled lane state
at $\beta = 5.0$ and $D=0.01$.
In order to highlight the effect of clogging on the transient time,
in our measurement we include only
lanes that contain a particle density equivalent to at least
80\% of the nominal number of particles per lane, $N/8$.
For each dense lane, the mean velocity evolves approximately
exponentially from zero to an almost fully polarized steady state flowing
in either the positive or negative $x$ direction.
The mean transient time
$\tau$ is then calculated as the average of the time constants obtained from exponential fits of the mean velocity data of each lane over 50 independent realizations.
For $N_{\text{p}, x} \geq 25$,
the reduction in the vertical mobility of the particles can cause
temporarily jammed states to form
in the lanes, so the time needed for the system to reach a steady state,
defined as having a lane collectively move to the left or right, increases.
An interesting effect is that as the particle density is increased
by increasing $N$,
more collisions between particles occur,
and this allows the system to reach a steady state faster.
This is the opposite of the effect that might have been expected, in
which 
higher densities of particles could increase the jamming probability. 
A similar behavior appears when we hold $N$ fixed and increase
$\beta$, as shown in Fig.~\ref{fig.7}(b) for a system with $N=528$.
At larger values of $\beta$, the lanes are better able to overcome transient jamming due to the high angular mobility of the particles. This suggests that enhancing the collective alignment of the particles in the lane, either by increasing $\beta$ or increasing the number of particles, reduces jamming, leading to smoother flows and shorter transient times.

\begin{figure}[!t]
\includegraphics[width= \linewidth]{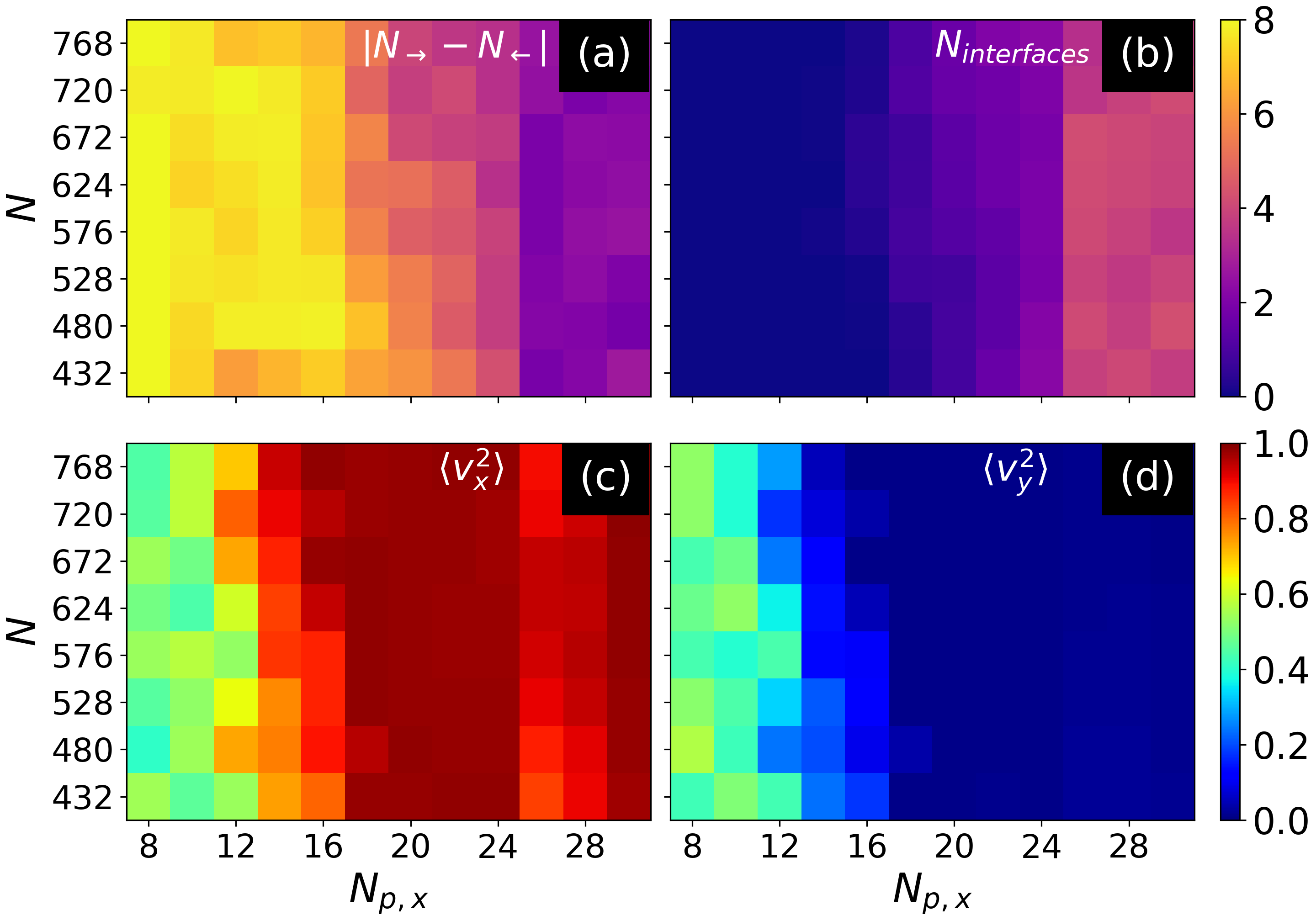}
\caption{
  Heat maps as a function of $N$ vs $N_{\text{p}, x}$.
  (a) The net difference between the number of right-moving lanes and left-moving lanes, $|N_\rightarrow - N_\leftarrow|$.
  (b) The number of velocity interfaces between neighboring lanes.
  (c) and (d) Mean square values of the $x$ and $y$ components of the particle
  velocities, $\langle v_{x}\rangle ^2$ and  $\langle v_{y}\rangle ^2$, respectively.
  When $N_{\text{p}, x}<15$,
  the system organizes into a globally polarized state with no velocity interfaces. In this state, both velocity components are finite, meaning that the particles can move either vertically or horizontally. All measures are averaged over 60 realizations of the random noise. 
  The results of these measurements were used to construct the schematic phase diagram in Fig.~\ref{fig.6}(d).
}
\label{fig.8}
\end{figure}

In Fig.~\ref{fig.8}, we plot heat maps showing how the various quantities change as a function of $N$ versus $N_{\text{p}, x}$.
Figure~\ref{fig.8}(a) shows the net difference between the number of right-moving lanes and left-moving lanes, $|N_\rightarrow-N_\leftarrow|$,
while in Fig.~\ref{fig.8}(b) we illustrate the number of velocity interfaces between neighboring lanes.
When $N_{\text{p}, x}<15$, $|N_\rightarrow-N_\leftarrow|\simeq8$ and the number of velocity interfaces is close to zero, indicating that in the quasi-isotropic phase, all of the particles move in the same direction and the system is globally polarized.
For $N_{\text{p}, x}=15$ to $25$ in the coupled lane state, we find a smaller value of $|N_\rightarrow-N_\leftarrow|\simeq5$ since some of the lanes are now moving in opposite directions.
For $N_{\text{p}, x}>25$, $|N_\rightarrow-N_\leftarrow|\approx 0$ and there are almost equal numbers of lanes
moving in each direction.
Similarly, when $N_{\text{p}, x}<16$ there are no velocity interfaces, for $16\leq N_{\text{p}, x}\leq 25$
there are two or three velocity interfaces, and for $N_{\text{p}, x}\geq 26$, there are multiple velocity interfaces.
In Fig.~\ref{fig.8}(c,d) we plot $\langle V_{x} \rangle ^2$ and $\langle V_{y} \rangle^2$, respectively. Both
velocity components are finite for $N_{\text{p}, x}<15$ in the quasi-isotropic phase.
For large values of $N_{\text{p}, x}$ the system shows 1D velocity polarization and the $y$-component velocity becomes nearly zero in the coupled and decoupled lane states.
We find that the boundaries between the quasi-isotropic, coupled, and decoupled lane states
are nearly insensitive to changes in $N$.
The features in the heat maps of Fig.~\ref{fig.8} were used to construct
the schematic phase diagram in Fig.~\ref{fig.6}(d).

\section{Summary}

In summary, we have numerically examined active matter particles with
both steric and alignment interactions in the presence of square and anisotropic obstacle substrates. In the
absence of a substrate, this system can form polarized or flocking states where the particles align their velocities in the
same direction at a critical alignment parameter.
In the presence of a square array, the location of the
alignment transition is not modified,
but for dense enough arrays,
the polarized velocity becomes
locked to one of the four symmetry directions of the substrate.
When we make the obstacle array increasingly anisotropic by
increasing the density of obstacles along the $x$ direction only,
in the polarized phase
we find a crossover from a quasi-isotropic state with global velocity polarization in both the $x$ and $y$ directions to a one-dimensional state in which
the velocity only polarizes along the $x$ direction.
For intermediate anisotropy, the particles can jump from lane to lane, and the system forms a coupled laned state containing
densely packed lanes of particles with the same polarizations
that can coexist with other more distant lanes that are oppositely polarized.
When the anisotropy is large enough, the particle hopping from lane to lane
is suppressed,
and the system forms a decoupled lane state
in which each lane forms a polarized state independent of the
polarization of neighboring lanes. Our results
can be generalized to a wide range of active matter or
flocking models with alignment interactions and could be
used to devise methods of controlling the direction of polarization.

\bigskip

\begin{acknowledgments}

This work was financed in part by Coordenação de Aperfeiçoamento de Pessoal de Nível Superior - Brasil (CAPES), Finance Code 001, Conselho Nacional de Desenvolvimento Científico e Tecnológico - Brasil (CNPq), Grant No.~312240/2021-0, and Fundação de Amparo à Ciência e Tecnologia do Estado de Pernambuco (FACEPE), Grant Number APQ-1129-1.05/24.
We gratefully acknowledge the support of the U.S. Department of
Energy through the LANL/LDRD program for this work.
This work was supported by the US Department of Energy through
the Los Alamos National Laboratory.  Los Alamos National Laboratory is
operated by Triad National Security, LLC, for the National Nuclear Security
Administration of the U. S. Department of Energy (Contract No. 892333218NCA000001).

\end{acknowledgments}

\bibliography{ActiveMatter.bib, Add.bib}

\begin{thebibliography}{42}%
\makeatletter
\providecommand \@ifxundefined [1]{%
 \@ifx{#1\undefined}
}%
\providecommand \@ifnum [1]{%
 \ifnum #1\expandafter \@firstoftwo
 \else \expandafter \@secondoftwo
 \fi
}%
\providecommand \@ifx [1]{%
 \ifx #1\expandafter \@firstoftwo
 \else \expandafter \@secondoftwo
 \fi
}%
\providecommand \natexlab [1]{#1}%
\providecommand \enquote  [1]{``#1''}%
\providecommand \bibnamefont  [1]{#1}%
\providecommand \bibfnamefont [1]{#1}%
\providecommand \citenamefont [1]{#1}%
\providecommand \href@noop [0]{\@secondoftwo}%
\providecommand \href [0]{\begingroup \@sanitize@url \@href}%
\providecommand \@href[1]{\@@startlink{#1}\@@href}%
\providecommand \@@href[1]{\endgroup#1\@@endlink}%
\providecommand \@sanitize@url [0]{\catcode `\\12\catcode `\$12\catcode `\&12\catcode `\#12\catcode `\^12\catcode `\_12\catcode `\%12\relax}%
\providecommand \@@startlink[1]{}%
\providecommand \@@endlink[0]{}%
\providecommand \url  [0]{\begingroup\@sanitize@url \@url }%
\providecommand \@url [1]{\endgroup\@href {#1}{\urlprefix }}%
\providecommand \urlprefix  [0]{URL }%
\providecommand \Eprint [0]{\href }%
\providecommand \doibase [0]{https://doi.org/}%
\providecommand \selectlanguage [0]{\@gobble}%
\providecommand \bibinfo  [0]{\@secondoftwo}%
\providecommand \bibfield  [0]{\@secondoftwo}%
\providecommand \translation [1]{[#1]}%
\providecommand \BibitemOpen [0]{}%
\providecommand \bibitemStop [0]{}%
\providecommand \bibitemNoStop [0]{.\EOS\space}%
\providecommand \EOS [0]{\spacefactor3000\relax}%
\providecommand \BibitemShut  [1]{\csname bibitem#1\endcsname}%
\let\auto@bib@innerbib\@empty
\bibitem [{\citenamefont {Marchetti}\ \emph {et~al.}(2013)\citenamefont {Marchetti}, \citenamefont {Joanny}, \citenamefont {Ramaswamy}, \citenamefont {Liverpool}, \citenamefont {Prost}, \citenamefont {Rao},\ and\ \citenamefont {Simha}}]{Marchetti2013review}%
  \BibitemOpen
  \bibfield  {author} {\bibinfo {author} {\bibfnamefont {M.~C.}\ \bibnamefont {Marchetti}}, \bibinfo {author} {\bibfnamefont {J.~F.}\ \bibnamefont {Joanny}}, \bibinfo {author} {\bibfnamefont {S.}~\bibnamefont {Ramaswamy}}, \bibinfo {author} {\bibfnamefont {T.~B.}\ \bibnamefont {Liverpool}}, \bibinfo {author} {\bibfnamefont {J.}~\bibnamefont {Prost}}, \bibinfo {author} {\bibfnamefont {M.}~\bibnamefont {Rao}},\ and\ \bibinfo {author} {\bibfnamefont {R.~A.}\ \bibnamefont {Simha}},\ }\bibfield  {title} {\bibinfo {title} {Hydrodynamics of soft active matter},\ }\href {https://doi.org/10.1103/RevModPhys.85.1143} {\bibfield  {journal} {\bibinfo  {journal} {Rev. Mod. Phys.}\ }\textbf {\bibinfo {volume} {85}},\ \bibinfo {pages} {1143} (\bibinfo {year} {2013})}\BibitemShut {NoStop}%
\bibitem [{\citenamefont {Bechinger}\ \emph {et~al.}(2016)\citenamefont {Bechinger}, \citenamefont {Di~Leonardo}, \citenamefont {L\"owen}, \citenamefont {Reichhardt}, \citenamefont {Volpe},\ and\ \citenamefont {Volpe}}]{Bechinger2016}%
  \BibitemOpen
  \bibfield  {author} {\bibinfo {author} {\bibfnamefont {C.}~\bibnamefont {Bechinger}}, \bibinfo {author} {\bibfnamefont {R.}~\bibnamefont {Di~Leonardo}}, \bibinfo {author} {\bibfnamefont {H.}~\bibnamefont {L\"owen}}, \bibinfo {author} {\bibfnamefont {C.}~\bibnamefont {Reichhardt}}, \bibinfo {author} {\bibfnamefont {G.}~\bibnamefont {Volpe}},\ and\ \bibinfo {author} {\bibfnamefont {G.}~\bibnamefont {Volpe}},\ }\bibfield  {title} {\bibinfo {title} {Active particles in complex and crowded environments},\ }\href {https://doi.org/10.1103/RevModPhys.88.045006} {\bibfield  {journal} {\bibinfo  {journal} {Rev. Mod. Phys.}\ }\textbf {\bibinfo {volume} {88}},\ \bibinfo {pages} {045006} (\bibinfo {year} {2016})}\BibitemShut {NoStop}%
\bibitem [{\citenamefont {Fily}\ and\ \citenamefont {Marchetti}(2012)}]{Fily_2012}%
  \BibitemOpen
  \bibfield  {author} {\bibinfo {author} {\bibfnamefont {Y.}~\bibnamefont {Fily}}\ and\ \bibinfo {author} {\bibfnamefont {M.~C.}\ \bibnamefont {Marchetti}},\ }\bibfield  {title} {\bibinfo {title} {Athermal phase separation of self-propelled particles with no alignment},\ }\href {https://doi.org/10.1103/physrevlett.108.235702} {\bibfield  {journal} {\bibinfo  {journal} {Physical Review Letters}\ }\textbf {\bibinfo {volume} {108}},\ \bibinfo {pages} {235702} (\bibinfo {year} {2012})}\BibitemShut {NoStop}%
\bibitem [{\citenamefont {Redner}\ \emph {et~al.}(2013)\citenamefont {Redner}, \citenamefont {Hagan},\ and\ \citenamefont {Baskaran}}]{Redner_2013}%
  \BibitemOpen
  \bibfield  {author} {\bibinfo {author} {\bibfnamefont {G.~S.}\ \bibnamefont {Redner}}, \bibinfo {author} {\bibfnamefont {M.~F.}\ \bibnamefont {Hagan}},\ and\ \bibinfo {author} {\bibfnamefont {A.}~\bibnamefont {Baskaran}},\ }\bibfield  {title} {\bibinfo {title} {Structure and dynamics of a phase-separating active colloidal fluid},\ }\href {https://doi.org/10.1016/j.bpj.2012.11.3534} {\bibfield  {journal} {\bibinfo  {journal} {Biophysical Journal}\ }\textbf {\bibinfo {volume} {104}},\ \bibinfo {pages} {640a} (\bibinfo {year} {2013})}\BibitemShut {NoStop}%
\bibitem [{\citenamefont {Palacci}\ \emph {et~al.}(2013)\citenamefont {Palacci}, \citenamefont {Sacanna}, \citenamefont {Steinberg}, \citenamefont {Pine},\ and\ \citenamefont {Chaikin}}]{Palacci13}%
  \BibitemOpen
  \bibfield  {author} {\bibinfo {author} {\bibfnamefont {J.}~\bibnamefont {Palacci}}, \bibinfo {author} {\bibfnamefont {S.}~\bibnamefont {Sacanna}}, \bibinfo {author} {\bibfnamefont {A.~P.}\ \bibnamefont {Steinberg}}, \bibinfo {author} {\bibfnamefont {D.~J.}\ \bibnamefont {Pine}},\ and\ \bibinfo {author} {\bibfnamefont {P.~M.}\ \bibnamefont {Chaikin}},\ }\bibfield  {title} {\bibinfo {title} {Living crystals of light-activated colloidal surfers},\ }\href {https://doi.org/10.1126/science.1230020} {\bibfield  {journal} {\bibinfo  {journal} {Science}\ }\textbf {\bibinfo {volume} {339}},\ \bibinfo {pages} {936} (\bibinfo {year} {2013})}\BibitemShut {NoStop}%
\bibitem [{\citenamefont {Cates}\ and\ \citenamefont {Tailleur}(2015)}]{Cates_2015}%
  \BibitemOpen
  \bibfield  {author} {\bibinfo {author} {\bibfnamefont {M.~E.}\ \bibnamefont {Cates}}\ and\ \bibinfo {author} {\bibfnamefont {J.}~\bibnamefont {Tailleur}},\ }\bibfield  {title} {\bibinfo {title} {Motility-induced phase separation},\ }\href {https://doi.org/10.1146/annurev-conmatphys-031214-014710} {\bibfield  {journal} {\bibinfo  {journal} {Annual Review of Condensed Matter Physics}\ }\textbf {\bibinfo {volume} {6}},\ \bibinfo {pages} {219–244} (\bibinfo {year} {2015})}\BibitemShut {NoStop}%
\bibitem [{\citenamefont {Volpe}\ \emph {et~al.}(2011{\natexlab{a}})\citenamefont {Volpe}, \citenamefont {Buttinoni}, \citenamefont {Vogt}, \citenamefont {K{\" u}mmerer},\ and\ \citenamefont {Bechinger}}]{Volpe11}%
  \BibitemOpen
  \bibfield  {author} {\bibinfo {author} {\bibfnamefont {G.}~\bibnamefont {Volpe}}, \bibinfo {author} {\bibfnamefont {I.}~\bibnamefont {Buttinoni}}, \bibinfo {author} {\bibfnamefont {D.}~\bibnamefont {Vogt}}, \bibinfo {author} {\bibfnamefont {H.-J.}\ \bibnamefont {K{\" u}mmerer}},\ and\ \bibinfo {author} {\bibfnamefont {C.}~\bibnamefont {Bechinger}},\ }\bibfield  {title} {\bibinfo {title} {Microswimmers in patterned environments},\ }\href {https://doi.org/10.1039/c1sm05960b} {\bibfield  {journal} {\bibinfo  {journal} {Soft Matter}\ }\textbf {\bibinfo {volume} {7}},\ \bibinfo {pages} {8810} (\bibinfo {year} {2011}{\natexlab{a}})}\BibitemShut {NoStop}%
\bibitem [{\citenamefont {Ribeiro}\ \emph {et~al.}(2020)\citenamefont {Ribeiro}, \citenamefont {Ferreira},\ and\ \citenamefont {Potiguar}}]{Ribeiro20}%
  \BibitemOpen
  \bibfield  {author} {\bibinfo {author} {\bibfnamefont {H.~E.}\ \bibnamefont {Ribeiro}}, \bibinfo {author} {\bibfnamefont {W.~P.}\ \bibnamefont {Ferreira}},\ and\ \bibinfo {author} {\bibfnamefont {F.~Q.}\ \bibnamefont {Potiguar}},\ }\bibfield  {title} {\bibinfo {title} {Trapping and sorting of active matter in a periodic background potential},\ }\href {https://doi.org/10.1103/PhysRevE.101.032126} {\bibfield  {journal} {\bibinfo  {journal} {Phys. Rev. E}\ }\textbf {\bibinfo {volume} {101}},\ \bibinfo {pages} {032126} (\bibinfo {year} {2020})}\BibitemShut {NoStop}%
\bibitem [{\citenamefont {Reichhardt}\ and\ \citenamefont {Reichhardt}(2021)}]{Reichhardt21-clogging}%
  \BibitemOpen
  \bibfield  {author} {\bibinfo {author} {\bibfnamefont {C.}~\bibnamefont {Reichhardt}}\ and\ \bibinfo {author} {\bibfnamefont {C.}~\bibnamefont {Reichhardt}},\ }\bibfield  {title} {\bibinfo {title} {Clogging, dynamics, and reentrant fluid for active matter on periodic substrates},\ }\href@noop {} {\bibfield  {journal} {\bibinfo  {journal} {Physical Review E}\ }\textbf {\bibinfo {volume} {103}},\ \bibinfo {pages} {062603} (\bibinfo {year} {2021})}\BibitemShut {NoStop}%
\bibitem [{\citenamefont {Kjeldbjerg}\ and\ \citenamefont {Brady}(2022)}]{Kjeldbjerg22}%
  \BibitemOpen
  \bibfield  {author} {\bibinfo {author} {\bibfnamefont {C.~M.}\ \bibnamefont {Kjeldbjerg}}\ and\ \bibinfo {author} {\bibfnamefont {J.~F.}\ \bibnamefont {Brady}},\ }\bibfield  {title} {\bibinfo {title} {Partitioning of active particles into porous media},\ }\href@noop {} {\bibfield  {journal} {\bibinfo  {journal} {Soft Matter}\ }\textbf {\bibinfo {volume} {18}},\ \bibinfo {pages} {2757} (\bibinfo {year} {2022})}\BibitemShut {NoStop}%
\bibitem [{\citenamefont {Modica}\ \emph {et~al.}(2023)\citenamefont {Modica}, \citenamefont {Omar},\ and\ \citenamefont {Takatori}}]{Modica23}%
  \BibitemOpen
  \bibfield  {author} {\bibinfo {author} {\bibfnamefont {K.~J.}\ \bibnamefont {Modica}}, \bibinfo {author} {\bibfnamefont {A.~K.}\ \bibnamefont {Omar}},\ and\ \bibinfo {author} {\bibfnamefont {S.~C.}\ \bibnamefont {Takatori}},\ }\bibfield  {title} {\bibinfo {title} {Boundary design regulates the diffusion of active matter in heterogeneous environments},\ }\href@noop {} {\bibfield  {journal} {\bibinfo  {journal} {Soft Matter}\ }\textbf {\bibinfo {volume} {19}},\ \bibinfo {pages} {1890} (\bibinfo {year} {2023})}\BibitemShut {NoStop}%
\bibitem [{\citenamefont {Reichhardt}\ and\ \citenamefont {Reichhardt}(2023)}]{reichhardt2023pattern}%
  \BibitemOpen
  \bibfield  {author} {\bibinfo {author} {\bibfnamefont {C.}~\bibnamefont {Reichhardt}}\ and\ \bibinfo {author} {\bibfnamefont {C.}~\bibnamefont {Reichhardt}},\ }\bibfield  {title} {\bibinfo {title} {Pattern formation and transport for externally driven active matter on periodic substrates (a)},\ }\href@noop {} {\bibfield  {journal} {\bibinfo  {journal} {Europhysics Letters}\ }\textbf {\bibinfo {volume} {142}},\ \bibinfo {pages} {37001} (\bibinfo {year} {2023})}\BibitemShut {NoStop}%
\bibitem [{\citenamefont {Chan}\ \emph {et~al.}(2024)\citenamefont {Chan}, \citenamefont {Wu}, \citenamefont {Qiao}, \citenamefont {Fong}, \citenamefont {Yang}, \citenamefont {Han},\ and\ \citenamefont {Zhang}}]{Chan24}%
  \BibitemOpen
  \bibfield  {author} {\bibinfo {author} {\bibfnamefont {C.~W.}\ \bibnamefont {Chan}}, \bibinfo {author} {\bibfnamefont {D.}~\bibnamefont {Wu}}, \bibinfo {author} {\bibfnamefont {K.}~\bibnamefont {Qiao}}, \bibinfo {author} {\bibfnamefont {K.~L.}\ \bibnamefont {Fong}}, \bibinfo {author} {\bibfnamefont {Z.}~\bibnamefont {Yang}}, \bibinfo {author} {\bibfnamefont {Y.}~\bibnamefont {Han}},\ and\ \bibinfo {author} {\bibfnamefont {R.}~\bibnamefont {Zhang}},\ }\bibfield  {title} {\bibinfo {title} {Chiral active particles are sensitive reporters to environmental geometry},\ }\href@noop {} {\bibfield  {journal} {\bibinfo  {journal} {Nature Communications}\ }\textbf {\bibinfo {volume} {15}},\ \bibinfo {pages} {1406} (\bibinfo {year} {2024})}\BibitemShut {NoStop}%
\bibitem [{\citenamefont {Caprini}\ \emph {et~al.}(2020{\natexlab{a}})\citenamefont {Caprini}, \citenamefont {Cecconi}, \citenamefont {Maggi},\ and\ \citenamefont {Marini Bettolo~Marconi}}]{caprini2020activity}%
  \BibitemOpen
  \bibfield  {author} {\bibinfo {author} {\bibfnamefont {L.}~\bibnamefont {Caprini}}, \bibinfo {author} {\bibfnamefont {F.}~\bibnamefont {Cecconi}}, \bibinfo {author} {\bibfnamefont {C.}~\bibnamefont {Maggi}},\ and\ \bibinfo {author} {\bibfnamefont {U.}~\bibnamefont {Marini Bettolo~Marconi}},\ }\bibfield  {title} {\bibinfo {title} {Activity-controlled clogging and unclogging of microchannels},\ }\href@noop {} {\bibfield  {journal} {\bibinfo  {journal} {Physical Review Research}\ }\textbf {\bibinfo {volume} {2}},\ \bibinfo {pages} {043359} (\bibinfo {year} {2020}{\natexlab{a}})}\BibitemShut {NoStop}%
\bibitem [{\citenamefont {Brun-Cosme-Bruny}\ \emph {et~al.}(2020)\citenamefont {Brun-Cosme-Bruny}, \citenamefont {F\"ortsch}, \citenamefont {Zimmermann}, \citenamefont {Bertin}, \citenamefont {Peyla},\ and\ \citenamefont {Rafa\"{\i}}}]{BrunCosmeBruny20}%
  \BibitemOpen
  \bibfield  {author} {\bibinfo {author} {\bibfnamefont {M.}~\bibnamefont {Brun-Cosme-Bruny}}, \bibinfo {author} {\bibfnamefont {A.}~\bibnamefont {F\"ortsch}}, \bibinfo {author} {\bibfnamefont {W.}~\bibnamefont {Zimmermann}}, \bibinfo {author} {\bibfnamefont {E.}~\bibnamefont {Bertin}}, \bibinfo {author} {\bibfnamefont {P.}~\bibnamefont {Peyla}},\ and\ \bibinfo {author} {\bibfnamefont {S.}~\bibnamefont {Rafa\"{\i}}},\ }\bibfield  {title} {\bibinfo {title} {Deflection of phototactic microswimmers through obstacle arrays},\ }\href {https://doi.org/10.1103/PhysRevFluids.5.093302} {\bibfield  {journal} {\bibinfo  {journal} {Phys. Rev. Fluids}\ }\textbf {\bibinfo {volume} {5}},\ \bibinfo {pages} {093302} (\bibinfo {year} {2020})}\BibitemShut {NoStop}%
\bibitem [{\citenamefont {Reichhardt}\ and\ \citenamefont {Reichhardt}(2020)}]{Reichhardt20}%
  \BibitemOpen
  \bibfield  {author} {\bibinfo {author} {\bibfnamefont {C.}~\bibnamefont {Reichhardt}}\ and\ \bibinfo {author} {\bibfnamefont {C.~J.~O.}\ \bibnamefont {Reichhardt}},\ }\bibfield  {title} {\bibinfo {title} {Directional locking effects for active matter particles coupled to a periodic substrate},\ }\href {https://doi.org/10.1103/PhysRevE.102.042616} {\bibfield  {journal} {\bibinfo  {journal} {Phys. Rev. E}\ }\textbf {\bibinfo {volume} {102}},\ \bibinfo {pages} {042616} (\bibinfo {year} {2020})}\BibitemShut {NoStop}%
\bibitem [{\citenamefont {Nabil}\ \emph {et~al.}(2022)\citenamefont {Nabil}, \citenamefont {Frankowski}, \citenamefont {Orosa}, \citenamefont {Fuller},\ and\ \citenamefont {Nourhani}}]{Nabil22}%
  \BibitemOpen
  \bibfield  {author} {\bibinfo {author} {\bibfnamefont {M.}~\bibnamefont {Nabil}}, \bibinfo {author} {\bibfnamefont {A.}~\bibnamefont {Frankowski}}, \bibinfo {author} {\bibfnamefont {A.}~\bibnamefont {Orosa}}, \bibinfo {author} {\bibfnamefont {A.}~\bibnamefont {Fuller}},\ and\ \bibinfo {author} {\bibfnamefont {A.}~\bibnamefont {Nourhani}},\ }\bibfield  {title} {\bibinfo {title} {Modulating drift dynamics of circle swimmers by periodic potentials},\ }\href@noop {} {\bibfield  {journal} {\bibinfo  {journal} {Physical Review E}\ }\textbf {\bibinfo {volume} {105}},\ \bibinfo {pages} {054610} (\bibinfo {year} {2022})}\BibitemShut {NoStop}%
\bibitem [{\citenamefont {Pattanayak}\ \emph {et~al.}(2019)\citenamefont {Pattanayak}, \citenamefont {Das}, \citenamefont {Kumar},\ and\ \citenamefont {Mishra}}]{pattanayak2019enhanced}%
  \BibitemOpen
  \bibfield  {author} {\bibinfo {author} {\bibfnamefont {S.}~\bibnamefont {Pattanayak}}, \bibinfo {author} {\bibfnamefont {R.}~\bibnamefont {Das}}, \bibinfo {author} {\bibfnamefont {M.}~\bibnamefont {Kumar}},\ and\ \bibinfo {author} {\bibfnamefont {S.}~\bibnamefont {Mishra}},\ }\bibfield  {title} {\bibinfo {title} {Enhanced dynamics of active brownian particles in periodic obstacle arrays and corrugated channels},\ }\href@noop {} {\bibfield  {journal} {\bibinfo  {journal} {The European Physical Journal E}\ }\textbf {\bibinfo {volume} {42}},\ \bibinfo {pages} {1} (\bibinfo {year} {2019})}\BibitemShut {NoStop}%
\bibitem [{\citenamefont {Fazelzadeh}\ \emph {et~al.}(2023)\citenamefont {Fazelzadeh}, \citenamefont {Di}, \citenamefont {Irani}, \citenamefont {Mokhtari},\ and\ \citenamefont {Jabbari-Farouji}}]{fazelzadeh2023active}%
  \BibitemOpen
  \bibfield  {author} {\bibinfo {author} {\bibfnamefont {M.}~\bibnamefont {Fazelzadeh}}, \bibinfo {author} {\bibfnamefont {Q.}~\bibnamefont {Di}}, \bibinfo {author} {\bibfnamefont {E.}~\bibnamefont {Irani}}, \bibinfo {author} {\bibfnamefont {Z.}~\bibnamefont {Mokhtari}},\ and\ \bibinfo {author} {\bibfnamefont {S.}~\bibnamefont {Jabbari-Farouji}},\ }\bibfield  {title} {\bibinfo {title} {Active motion of tangentially driven polymers in periodic array of obstacles},\ }\href@noop {} {\bibfield  {journal} {\bibinfo  {journal} {The Journal of Chemical Physics}\ }\textbf {\bibinfo {volume} {159}} (\bibinfo {year} {2023})}\BibitemShut {NoStop}%
\bibitem [{\citenamefont {Nayak}\ \emph {et~al.}(2023)\citenamefont {Nayak}, \citenamefont {Das}, \citenamefont {Bag}, \citenamefont {Debnath},\ and\ \citenamefont {Ghosh}}]{nayak2023driven}%
  \BibitemOpen
  \bibfield  {author} {\bibinfo {author} {\bibfnamefont {S.}~\bibnamefont {Nayak}}, \bibinfo {author} {\bibfnamefont {S.}~\bibnamefont {Das}}, \bibinfo {author} {\bibfnamefont {P.}~\bibnamefont {Bag}}, \bibinfo {author} {\bibfnamefont {T.}~\bibnamefont {Debnath}},\ and\ \bibinfo {author} {\bibfnamefont {P.~K.}\ \bibnamefont {Ghosh}},\ }\bibfield  {title} {\bibinfo {title} {Driven transport of active particles through arrays of symmetric obstacles},\ }\href@noop {} {\bibfield  {journal} {\bibinfo  {journal} {The Journal of Chemical Physics}\ }\textbf {\bibinfo {volume} {159}} (\bibinfo {year} {2023})}\BibitemShut {NoStop}%
\bibitem [{\citenamefont {Vent{\'e}jou}\ \emph {et~al.}(2024)\citenamefont {Vent{\'e}jou}, \citenamefont {Magniez-Papillon}, \citenamefont {Bertin}, \citenamefont {Peyla},\ and\ \citenamefont {Dupont}}]{ventejou2024behavioral}%
  \BibitemOpen
  \bibfield  {author} {\bibinfo {author} {\bibfnamefont {B.}~\bibnamefont {Vent{\'e}jou}}, \bibinfo {author} {\bibfnamefont {I.}~\bibnamefont {Magniez-Papillon}}, \bibinfo {author} {\bibfnamefont {E.}~\bibnamefont {Bertin}}, \bibinfo {author} {\bibfnamefont {P.}~\bibnamefont {Peyla}},\ and\ \bibinfo {author} {\bibfnamefont {A.}~\bibnamefont {Dupont}},\ }\bibfield  {title} {\bibinfo {title} {Behavioral transition of a fish school in a crowded environment},\ }\href@noop {} {\bibfield  {journal} {\bibinfo  {journal} {Physical Review E}\ }\textbf {\bibinfo {volume} {109}},\ \bibinfo {pages} {064403} (\bibinfo {year} {2024})}\BibitemShut {NoStop}%
\bibitem [{\citenamefont {Bowick}\ \emph {et~al.}(2022)\citenamefont {Bowick}, \citenamefont {Fakhri}, \citenamefont {Marchetti},\ and\ \citenamefont {Ramaswamy}}]{bowick2022symmetry}%
  \BibitemOpen
  \bibfield  {author} {\bibinfo {author} {\bibfnamefont {M.~J.}\ \bibnamefont {Bowick}}, \bibinfo {author} {\bibfnamefont {N.}~\bibnamefont {Fakhri}}, \bibinfo {author} {\bibfnamefont {M.~C.}\ \bibnamefont {Marchetti}},\ and\ \bibinfo {author} {\bibfnamefont {S.}~\bibnamefont {Ramaswamy}},\ }\bibfield  {title} {\bibinfo {title} {Symmetry, thermodynamics, and topology in active matter},\ }\href@noop {} {\bibfield  {journal} {\bibinfo  {journal} {Physical Review X}\ }\textbf {\bibinfo {volume} {12}},\ \bibinfo {pages} {010501} (\bibinfo {year} {2022})}\BibitemShut {NoStop}%
\bibitem [{\citenamefont {Dauchot}\ and\ \citenamefont {D\'emery}(2019)}]{Dauchot2019}%
  \BibitemOpen
  \bibfield  {author} {\bibinfo {author} {\bibfnamefont {O.}~\bibnamefont {Dauchot}}\ and\ \bibinfo {author} {\bibfnamefont {V.}~\bibnamefont {D\'emery}},\ }\bibfield  {title} {\bibinfo {title} {Dynamics of a self-propelled particle in a harmonic trap},\ }\href {https://doi.org/10.1103/PhysRevLett.122.068002} {\bibfield  {journal} {\bibinfo  {journal} {Phys. Rev. Lett.}\ }\textbf {\bibinfo {volume} {122}},\ \bibinfo {pages} {068002} (\bibinfo {year} {2019})}\BibitemShut {NoStop}%
\bibitem [{\citenamefont {Damascena}\ \emph {et~al.}(2022)\citenamefont {Damascena}, \citenamefont {Cabral},\ and\ \citenamefont {Silva}}]{Rubens2022}%
  \BibitemOpen
  \bibfield  {author} {\bibinfo {author} {\bibfnamefont {R.~H.}\ \bibnamefont {Damascena}}, \bibinfo {author} {\bibfnamefont {L.~R.~E.}\ \bibnamefont {Cabral}},\ and\ \bibinfo {author} {\bibfnamefont {C.~C. d.~S.}\ \bibnamefont {Silva}},\ }\bibfield  {title} {\bibinfo {title} {Coexisting orbits and chaotic dynamics of a confined self-propelled particle},\ }\href {https://doi.org/10.1103/PhysRevE.105.064608} {\bibfield  {journal} {\bibinfo  {journal} {Phys. Rev. E}\ }\textbf {\bibinfo {volume} {105}},\ \bibinfo {pages} {064608} (\bibinfo {year} {2022})}\BibitemShut {NoStop}%
\bibitem [{\citenamefont {Damascena}\ and\ \citenamefont {de~Souza~Silva}(2023)}]{Rubens2023}%
  \BibitemOpen
  \bibfield  {author} {\bibinfo {author} {\bibfnamefont {R.~H.}\ \bibnamefont {Damascena}}\ and\ \bibinfo {author} {\bibfnamefont {C.~C.}\ \bibnamefont {de~Souza~Silva}},\ }\bibfield  {title} {\bibinfo {title} {Noise-induced escape of a self-propelled particle from metastable orbits},\ }\href {https://doi.org/10.1103/PhysRevE.108.044605} {\bibfield  {journal} {\bibinfo  {journal} {Phys. Rev. E}\ }\textbf {\bibinfo {volume} {108}},\ \bibinfo {pages} {044605} (\bibinfo {year} {2023})}\BibitemShut {NoStop}%
\bibitem [{\citenamefont {Giavazzi}\ \emph {et~al.}(2018)\citenamefont {Giavazzi}, \citenamefont {Paoluzzi}, \citenamefont {Macchi}, \citenamefont {Bi}, \citenamefont {Scita}, \citenamefont {Manning}, \citenamefont {Cerbino},\ and\ \citenamefont {Marchetti}}]{giavazzi2018flocking}%
  \BibitemOpen
  \bibfield  {author} {\bibinfo {author} {\bibfnamefont {F.}~\bibnamefont {Giavazzi}}, \bibinfo {author} {\bibfnamefont {M.}~\bibnamefont {Paoluzzi}}, \bibinfo {author} {\bibfnamefont {M.}~\bibnamefont {Macchi}}, \bibinfo {author} {\bibfnamefont {D.}~\bibnamefont {Bi}}, \bibinfo {author} {\bibfnamefont {G.}~\bibnamefont {Scita}}, \bibinfo {author} {\bibfnamefont {M.~L.}\ \bibnamefont {Manning}}, \bibinfo {author} {\bibfnamefont {R.}~\bibnamefont {Cerbino}},\ and\ \bibinfo {author} {\bibfnamefont {M.~C.}\ \bibnamefont {Marchetti}},\ }\bibfield  {title} {\bibinfo {title} {Flocking transitions in confluent tissues},\ }\href@noop {} {\bibfield  {journal} {\bibinfo  {journal} {Soft matter}\ }\textbf {\bibinfo {volume} {14}},\ \bibinfo {pages} {3471} (\bibinfo {year} {2018})}\BibitemShut {NoStop}%
\bibitem [{\citenamefont {Geyer}\ \emph {et~al.}(2019)\citenamefont {Geyer}, \citenamefont {Martin}, \citenamefont {Tailleur},\ and\ \citenamefont {Bartolo}}]{geyer2019freezing}%
  \BibitemOpen
  \bibfield  {author} {\bibinfo {author} {\bibfnamefont {D.}~\bibnamefont {Geyer}}, \bibinfo {author} {\bibfnamefont {D.}~\bibnamefont {Martin}}, \bibinfo {author} {\bibfnamefont {J.}~\bibnamefont {Tailleur}},\ and\ \bibinfo {author} {\bibfnamefont {D.}~\bibnamefont {Bartolo}},\ }\bibfield  {title} {\bibinfo {title} {Freezing a flock: Motility-induced phase separation in polar active liquids},\ }\href@noop {} {\bibfield  {journal} {\bibinfo  {journal} {Physical Review X}\ }\textbf {\bibinfo {volume} {9}},\ \bibinfo {pages} {031043} (\bibinfo {year} {2019})}\BibitemShut {NoStop}%
\bibitem [{\citenamefont {Caprini}\ \emph {et~al.}(2020{\natexlab{b}})\citenamefont {Caprini}, \citenamefont {Marini Bettolo~Marconi},\ and\ \citenamefont {Puglisi}}]{caprini2020spontaneous}%
  \BibitemOpen
  \bibfield  {author} {\bibinfo {author} {\bibfnamefont {L.}~\bibnamefont {Caprini}}, \bibinfo {author} {\bibfnamefont {U.}~\bibnamefont {Marini Bettolo~Marconi}},\ and\ \bibinfo {author} {\bibfnamefont {A.}~\bibnamefont {Puglisi}},\ }\bibfield  {title} {\bibinfo {title} {Spontaneous velocity alignment in motility-induced phase separation},\ }\href@noop {} {\bibfield  {journal} {\bibinfo  {journal} {Physical review letters}\ }\textbf {\bibinfo {volume} {124}},\ \bibinfo {pages} {078001} (\bibinfo {year} {2020}{\natexlab{b}})}\BibitemShut {NoStop}%
\bibitem [{\citenamefont {Fazli}\ and\ \citenamefont {Naji}(2021)}]{fazli2021active}%
  \BibitemOpen
  \bibfield  {author} {\bibinfo {author} {\bibfnamefont {Z.}~\bibnamefont {Fazli}}\ and\ \bibinfo {author} {\bibfnamefont {A.}~\bibnamefont {Naji}},\ }\bibfield  {title} {\bibinfo {title} {Active particles with polar alignment in ring-shaped confinement},\ }\href@noop {} {\bibfield  {journal} {\bibinfo  {journal} {Physical Review E}\ }\textbf {\bibinfo {volume} {103}},\ \bibinfo {pages} {022601} (\bibinfo {year} {2021})}\BibitemShut {NoStop}%
\bibitem [{\citenamefont {Canavello}\ \emph {et~al.}(2024)\citenamefont {Canavello}, \citenamefont {Damascena}, \citenamefont {Cabral},\ and\ \citenamefont {de~Souza~Silva}}]{Daniel2024}%
  \BibitemOpen
  \bibfield  {author} {\bibinfo {author} {\bibfnamefont {D.}~\bibnamefont {Canavello}}, \bibinfo {author} {\bibfnamefont {R.~H.}\ \bibnamefont {Damascena}}, \bibinfo {author} {\bibfnamefont {L.~R.~E.}\ \bibnamefont {Cabral}},\ and\ \bibinfo {author} {\bibfnamefont {C.~C.}\ \bibnamefont {de~Souza~Silva}},\ }\bibfield  {title} {\bibinfo {title} {Polar order{,} shear banding{,} and clustering in confined active matter},\ }\href {https://doi.org/10.1039/D3SM01721D} {\bibfield  {journal} {\bibinfo  {journal} {Soft Matter}\ }\textbf {\bibinfo {volume} {20}},\ \bibinfo {pages} {2310} (\bibinfo {year} {2024})}\BibitemShut {NoStop}%
\bibitem [{\citenamefont {Briand}\ \emph {et~al.}(2018)\citenamefont {Briand}, \citenamefont {Schindler},\ and\ \citenamefont {Dauchot}}]{briand2018spontaneously}%
  \BibitemOpen
  \bibfield  {author} {\bibinfo {author} {\bibfnamefont {G.}~\bibnamefont {Briand}}, \bibinfo {author} {\bibfnamefont {M.}~\bibnamefont {Schindler}},\ and\ \bibinfo {author} {\bibfnamefont {O.}~\bibnamefont {Dauchot}},\ }\bibfield  {title} {\bibinfo {title} {Spontaneously flowing crystal of self-propelled particles},\ }\href@noop {} {\bibfield  {journal} {\bibinfo  {journal} {Physical review letters}\ }\textbf {\bibinfo {volume} {120}},\ \bibinfo {pages} {208001} (\bibinfo {year} {2018})}\BibitemShut {NoStop}%
\bibitem [{\citenamefont {Baconnier}\ \emph {et~al.}(2022)\citenamefont {Baconnier}, \citenamefont {Shohat}, \citenamefont {L{\'o}pez}, \citenamefont {Coulais}, \citenamefont {D{\'e}mery}, \citenamefont {D{\"u}ring},\ and\ \citenamefont {Dauchot}}]{baconnier2022selective}%
  \BibitemOpen
  \bibfield  {author} {\bibinfo {author} {\bibfnamefont {P.}~\bibnamefont {Baconnier}}, \bibinfo {author} {\bibfnamefont {D.}~\bibnamefont {Shohat}}, \bibinfo {author} {\bibfnamefont {C.~H.}\ \bibnamefont {L{\'o}pez}}, \bibinfo {author} {\bibfnamefont {C.}~\bibnamefont {Coulais}}, \bibinfo {author} {\bibfnamefont {V.}~\bibnamefont {D{\'e}mery}}, \bibinfo {author} {\bibfnamefont {G.}~\bibnamefont {D{\"u}ring}},\ and\ \bibinfo {author} {\bibfnamefont {O.}~\bibnamefont {Dauchot}},\ }\bibfield  {title} {\bibinfo {title} {Selective and collective actuation in active solids},\ }\href@noop {} {\bibfield  {journal} {\bibinfo  {journal} {Nature Physics}\ }\textbf {\bibinfo {volume} {18}},\ \bibinfo {pages} {1234} (\bibinfo {year} {2022})}\BibitemShut {NoStop}%
\bibitem [{\citenamefont {Baconnier}\ \emph {et~al.}(2024)\citenamefont {Baconnier}, \citenamefont {Dauchot}, \citenamefont {D{\'e}mery}, \citenamefont {D{\"u}ring}, \citenamefont {Henkes}, \citenamefont {Huepe},\ and\ \citenamefont {Shee}}]{baconnier2024self}%
  \BibitemOpen
  \bibfield  {author} {\bibinfo {author} {\bibfnamefont {P.}~\bibnamefont {Baconnier}}, \bibinfo {author} {\bibfnamefont {O.}~\bibnamefont {Dauchot}}, \bibinfo {author} {\bibfnamefont {V.}~\bibnamefont {D{\'e}mery}}, \bibinfo {author} {\bibfnamefont {G.}~\bibnamefont {D{\"u}ring}}, \bibinfo {author} {\bibfnamefont {S.}~\bibnamefont {Henkes}}, \bibinfo {author} {\bibfnamefont {C.}~\bibnamefont {Huepe}},\ and\ \bibinfo {author} {\bibfnamefont {A.}~\bibnamefont {Shee}},\ }\bibfield  {title} {\bibinfo {title} {Self-aligning polar active matter},\ }\href@noop {} {\bibfield  {journal} {\bibinfo  {journal} {arXiv preprint arXiv:2403.10151}\ } (\bibinfo {year} {2024})}\BibitemShut {NoStop}%
\bibitem [{\citenamefont {Briand}\ and\ \citenamefont {Dauchot}(2016)}]{briand2016crystallization}%
  \BibitemOpen
  \bibfield  {author} {\bibinfo {author} {\bibfnamefont {G.}~\bibnamefont {Briand}}\ and\ \bibinfo {author} {\bibfnamefont {O.}~\bibnamefont {Dauchot}},\ }\bibfield  {title} {\bibinfo {title} {Crystallization of self-propelled hard discs},\ }\href@noop {} {\bibfield  {journal} {\bibinfo  {journal} {Physical review letters}\ }\textbf {\bibinfo {volume} {117}},\ \bibinfo {pages} {098004} (\bibinfo {year} {2016})}\BibitemShut {NoStop}%
\bibitem [{\citenamefont {Howse}\ \emph {et~al.}(2007)\citenamefont {Howse}, \citenamefont {Jones}, \citenamefont {Ryan}, \citenamefont {Gough}, \citenamefont {Vafabakhsh},\ and\ \citenamefont {Golestanian}}]{Howse_2007}%
  \BibitemOpen
  \bibfield  {author} {\bibinfo {author} {\bibfnamefont {J.~R.}\ \bibnamefont {Howse}}, \bibinfo {author} {\bibfnamefont {R.~A.~L.}\ \bibnamefont {Jones}}, \bibinfo {author} {\bibfnamefont {A.~J.}\ \bibnamefont {Ryan}}, \bibinfo {author} {\bibfnamefont {T.}~\bibnamefont {Gough}}, \bibinfo {author} {\bibfnamefont {R.}~\bibnamefont {Vafabakhsh}},\ and\ \bibinfo {author} {\bibfnamefont {R.}~\bibnamefont {Golestanian}},\ }\bibfield  {title} {\bibinfo {title} {Self-motile colloidal particles: From directed propulsion to random walk},\ }\href {https://doi.org/10.1103/physrevlett.99.048102} {\bibfield  {journal} {\bibinfo  {journal} {Physical Review Letters}\ }\textbf {\bibinfo {volume} {99}},\ \bibinfo {pages} {048102} (\bibinfo {year} {2007})}\BibitemShut {NoStop}%
\bibitem [{\citenamefont {Volpe}\ \emph {et~al.}(2011{\natexlab{b}})\citenamefont {Volpe}, \citenamefont {Buttinoni}, \citenamefont {Vogt}, \citenamefont {K{\"u}mmerer},\ and\ \citenamefont {Bechinger}}]{volpe2011}%
  \BibitemOpen
  \bibfield  {author} {\bibinfo {author} {\bibfnamefont {G.}~\bibnamefont {Volpe}}, \bibinfo {author} {\bibfnamefont {I.}~\bibnamefont {Buttinoni}}, \bibinfo {author} {\bibfnamefont {D.}~\bibnamefont {Vogt}}, \bibinfo {author} {\bibfnamefont {H.-J.}\ \bibnamefont {K{\"u}mmerer}},\ and\ \bibinfo {author} {\bibfnamefont {C.}~\bibnamefont {Bechinger}},\ }\bibfield  {title} {\bibinfo {title} {Microswimmers in patterned environments},\ }\href@noop {} {\bibfield  {journal} {\bibinfo  {journal} {Soft Matter}\ }\textbf {\bibinfo {volume} {7}},\ \bibinfo {pages} {8810} (\bibinfo {year} {2011}{\natexlab{b}})}\BibitemShut {NoStop}%
\bibitem [{\citenamefont {Drescher}\ \emph {et~al.}(2011)\citenamefont {Drescher}, \citenamefont {Dunkel}, \citenamefont {Cisneros}, \citenamefont {Ganguly},\ and\ \citenamefont {Goldstein}}]{Drescher2011}%
  \BibitemOpen
  \bibfield  {author} {\bibinfo {author} {\bibfnamefont {K.}~\bibnamefont {Drescher}}, \bibinfo {author} {\bibfnamefont {J.}~\bibnamefont {Dunkel}}, \bibinfo {author} {\bibfnamefont {L.~H.}\ \bibnamefont {Cisneros}}, \bibinfo {author} {\bibfnamefont {S.}~\bibnamefont {Ganguly}},\ and\ \bibinfo {author} {\bibfnamefont {R.~E.}\ \bibnamefont {Goldstein}},\ }\bibfield  {title} {\bibinfo {title} {Fluid dynamics and noise in bacterial cell–cell and cell–surface scattering},\ }\href {https://doi.org/10.1073/pnas.1019079108} {\bibfield  {journal} {\bibinfo  {journal} {Proceedings of the National Academy of Sciences}\ }\textbf {\bibinfo {volume} {108}},\ \bibinfo {pages} {10940} (\bibinfo {year} {2011})},\ \Eprint {https://arxiv.org/abs/https://www.pnas.org/doi/pdf/10.1073/pnas.1019079108} {https://www.pnas.org/doi/pdf/10.1073/pnas.1019079108} \BibitemShut {NoStop}%
\bibitem [{\citenamefont {Shimoyama}\ \emph {et~al.}(1996)\citenamefont {Shimoyama}, \citenamefont {Sugawara}, \citenamefont {Mizuguchi}, \citenamefont {Hayakawa},\ and\ \citenamefont {Sano}}]{Shimoyama1996}%
  \BibitemOpen
  \bibfield  {author} {\bibinfo {author} {\bibfnamefont {N.}~\bibnamefont {Shimoyama}}, \bibinfo {author} {\bibfnamefont {K.}~\bibnamefont {Sugawara}}, \bibinfo {author} {\bibfnamefont {T.}~\bibnamefont {Mizuguchi}}, \bibinfo {author} {\bibfnamefont {Y.}~\bibnamefont {Hayakawa}},\ and\ \bibinfo {author} {\bibfnamefont {M.}~\bibnamefont {Sano}},\ }\bibfield  {title} {\bibinfo {title} {Collective motion in a system of motile elements},\ }\href {https://doi.org/10.1103/PhysRevLett.76.3870} {\bibfield  {journal} {\bibinfo  {journal} {Phys. Rev. Lett.}\ }\textbf {\bibinfo {volume} {76}},\ \bibinfo {pages} {3870} (\bibinfo {year} {1996})}\BibitemShut {NoStop}%
\bibitem [{\citenamefont {Henkes}\ \emph {et~al.}(2011)\citenamefont {Henkes}, \citenamefont {Fily},\ and\ \citenamefont {Marchetti}}]{Henkes2011}%
  \BibitemOpen
  \bibfield  {author} {\bibinfo {author} {\bibfnamefont {S.}~\bibnamefont {Henkes}}, \bibinfo {author} {\bibfnamefont {Y.}~\bibnamefont {Fily}},\ and\ \bibinfo {author} {\bibfnamefont {M.~C.}\ \bibnamefont {Marchetti}},\ }\bibfield  {title} {\bibinfo {title} {Active jamming: Self-propelled soft particles at high density},\ }\href {https://doi.org/10.1103/PhysRevE.84.040301} {\bibfield  {journal} {\bibinfo  {journal} {Phys. Rev. E}\ }\textbf {\bibinfo {volume} {84}},\ \bibinfo {pages} {040301} (\bibinfo {year} {2011})}\BibitemShut {NoStop}%
\bibitem [{\citenamefont {Weber}\ \emph {et~al.}(2013)\citenamefont {Weber}, \citenamefont {Hanke}, \citenamefont {Deseigne}, \citenamefont {L\'eonard}, \citenamefont {Dauchot}, \citenamefont {Frey},\ and\ \citenamefont {Chat\'e}}]{Weber2013}%
  \BibitemOpen
  \bibfield  {author} {\bibinfo {author} {\bibfnamefont {C.~A.}\ \bibnamefont {Weber}}, \bibinfo {author} {\bibfnamefont {T.}~\bibnamefont {Hanke}}, \bibinfo {author} {\bibfnamefont {J.}~\bibnamefont {Deseigne}}, \bibinfo {author} {\bibfnamefont {S.}~\bibnamefont {L\'eonard}}, \bibinfo {author} {\bibfnamefont {O.}~\bibnamefont {Dauchot}}, \bibinfo {author} {\bibfnamefont {E.}~\bibnamefont {Frey}},\ and\ \bibinfo {author} {\bibfnamefont {H.}~\bibnamefont {Chat\'e}},\ }\bibfield  {title} {\bibinfo {title} {Long-range ordering of vibrated polar disks},\ }\href {https://doi.org/10.1103/PhysRevLett.110.208001} {\bibfield  {journal} {\bibinfo  {journal} {Phys. Rev. Lett.}\ }\textbf {\bibinfo {volume} {110}},\ \bibinfo {pages} {208001} (\bibinfo {year} {2013})}\BibitemShut {NoStop}%
\bibitem [{\citenamefont {Lam}\ \emph {et~al.}(2015)\citenamefont {Lam}, \citenamefont {Schindler},\ and\ \citenamefont {Dauchot}}]{Lam2015}%
  \BibitemOpen
  \bibfield  {author} {\bibinfo {author} {\bibfnamefont {K.-D. N.~T.}\ \bibnamefont {Lam}}, \bibinfo {author} {\bibfnamefont {M.}~\bibnamefont {Schindler}},\ and\ \bibinfo {author} {\bibfnamefont {O.}~\bibnamefont {Dauchot}},\ }\bibfield  {title} {\bibinfo {title} {Self-propelled hard disks: implicit alignment and transition to collective motion},\ }\href {https://doi.org/10.1088/1367-2630/17/11/113056} {\bibfield  {journal} {\bibinfo  {journal} {New Journal of Physics}\ }\textbf {\bibinfo {volume} {17}},\ \bibinfo {pages} {113056} (\bibinfo {year} {2015})}\BibitemShut {NoStop}%
\bibitem [{\citenamefont {Martin-Roca}\ \emph {et~al.}(2021)\citenamefont {Martin-Roca}, \citenamefont {Martinez}, \citenamefont {Alexander}, \citenamefont {Diez}, \citenamefont {Aarts}, \citenamefont {Alarcon}, \citenamefont {Ramírez},\ and\ \citenamefont {Valeriani}}]{Roca2021}%
  \BibitemOpen
  \bibfield  {author} {\bibinfo {author} {\bibfnamefont {J.}~\bibnamefont {Martin-Roca}}, \bibinfo {author} {\bibfnamefont {R.}~\bibnamefont {Martinez}}, \bibinfo {author} {\bibfnamefont {L.~C.}\ \bibnamefont {Alexander}}, \bibinfo {author} {\bibfnamefont {A.~L.}\ \bibnamefont {Diez}}, \bibinfo {author} {\bibfnamefont {D.~G. A.~L.}\ \bibnamefont {Aarts}}, \bibinfo {author} {\bibfnamefont {F.}~\bibnamefont {Alarcon}}, \bibinfo {author} {\bibfnamefont {J.}~\bibnamefont {Ramírez}},\ and\ \bibinfo {author} {\bibfnamefont {C.}~\bibnamefont {Valeriani}},\ }\bibfield  {title} {\bibinfo {title} {{Characterization of MIPS in a suspension of repulsive active Brownian particles through dynamical features}},\ }\href {https://doi.org/10.1063/5.0040141} {\bibfield  {journal} {\bibinfo  {journal} {The Journal of Chemical Physics}\ }\textbf {\bibinfo {volume} {154}},\ \bibinfo {pages} {164901} (\bibinfo {year} {2021})}\BibitemShut {NoStop}%
\end{thebibliography}%

\end{document}